\documentclass[a4paper,11pt]{article}
\pdfoutput=1

\usepackage{jheppub}
\usepackage[T1]{fontenc}
\usepackage{amssymb}
\usepackage{float}
\usepackage{amsmath}

\usepackage{graphicx}
\usepackage{tensor}
\usepackage{slashed}
\usepackage[utf8]{inputenc}
\usepackage{hyperref}

\newcommand{\td}{\text{d}}

\newcommand{\coh}{\text{coh}}

\begin{document}
\title{\boldmath A Complexity for Quantum Field Theory States and Application in Thermofield Double States}

\author{Run-Qiu Yang}

\emailAdd{aqiu@kias.re.kr}

\affiliation{Quantum Universe Center, Korea Institute for Advanced Study, Seoul 130-722, Korea}

\abstract{This paper defines a complexity between states in quantum field theory by introducing a Finsler structure based on ladder operators (the generalization of creation and annihilation operators). Two simple models are shown as examples to clarify the differences between complexity and other conceptions such as complexity of formation and entanglement entropy. When it is applied into thermofield double (TFD) states in $d$-dimensional conformal field theory, results show that the complexity density between them and corresponding vacuum states are finite and proportional to $T^{d-1}$, where $T$ is the temperature of TFD state. Especially, a proof is given to show that fidelity susceptibility of  a TFD state is equivalent to the complexity between it and corresponding vacuum state, which gives an explanation why they may share the same object in holographic duality. Some enlightenments to holographic conjectures of complexity are also discussed.
}

\maketitle
\flushbottom

\noindent

\section{Introduction}\label{Intro}
In recent years, the conceptions in quantum information theory are applying into the understanding about the high energy and gravity physics. This leads to some surprising discoveries about the connection of entanglement and geometry~\cite{VanRaamsdonk:2010pw,Ryu:2006bv,Maldacena:2013xja,Faulkner2014}.  Especially, the consideration about the some aspects in the  wormhole created by an Einstein-Rosen (ER) bridge~\cite{PhysRev.48.73} and a pair of maximally entangled black holes leads  Leonard Susskind and Juan Maldacena to propose a very interesting conjecture named ``EPR=ER''~\cite{Maldacena:2013xja}. Here EPR refers to quantum entanglement (EPR paradox). The deeper consideration about ``EPR=ER'' leads to a quantity named ``complexity'' and its holographic descriptions~\cite{Susskind:2014moa}.

In this study, they consider an eternal AdS black hole which is conjectured to dual to a thermofield double (TFD) state,
\begin{equation}\label{TFD1}
  |\text{TFD}\rangle:=Z^{-1/2}\sum_\alpha\exp[-E_\alpha/(2T)]|E_\alpha\rangle_L |E_\alpha\rangle_R \,.
\end{equation}
The states $|E_\alpha\rangle_L$ and $|E_\alpha\rangle_R$ are defined in the two copy CFTs at the two boundaries of the eternal AdS black hole and $T$ is the temperature. With the Hamiltonians $H_L$ and $H_R$ at the left and right dual CFTs, the time evolution of a TFD state is
\begin{equation}\label{timesate1}
  |\psi(t_L,t_R)\rangle:=e^{-i(t_LH_L+t_RH_R)}|\text{TFD}\rangle\,,
\end{equation}
which can be characterized by two codimension-two surfaces at the two boundaries of the AdS black hole with left time $t=t_L$ and right time $t=t_R$~\cite{Maldacena:2001kr,Brown:2015lvg}. There are two proposals to compute the complexity for $|\psi(t_L,t_R)\rangle$ holographically:\footnote{There are also some other holographic proposals for complexity, see Refs.~\cite{Alishahiha:2015rta,Ben-Ami:2016qex,Couch:2016exn} for examples. } CV(complexity=volume) conjecture~\cite{Susskind:2014rva,Stanford:2014jda,Alishahiha:2015rta} and CA(complexity= action) conjecture~\cite{Brown:2015bva,Brown:2015lvg}.

The CV conjecture~\cite{Stanford:2014jda,Alishahiha:2015rta} states that the complexity of $|\psi(t_L,t_R)\rangle$ is proportional to the maximal volume of the space-like codimension-one surfaces which connect the codimension-two time-like slices denoted by $t_L$ and $t_R$ at the two AdS boundaries, i.e.
\begin{equation}\label{CV}
  \mathcal{C}_V=\max_{\partial \Sigma=t_L\cup t_R}\left[\frac{V(\Sigma)}{G_N \ell}\right] \,,
\end{equation}
where $G_N$ is the Newton's constant. $\Sigma$ is the possible space-like codimension-one surface which connects $t_L$ and $t_R$. $\ell$ is a length scale associated with the bulk geometry such as horizon radius or AdS radius and so on.
This conjecture satisfies some  properties of the quantum complexity. However, there is an ambiguity coming from the choice of a length scale $\ell$. This unsatisfactory feature motivated the second conjecture: CA conjecture \cite{Brown:2015bva,Brown:2015lvg}. In this conjecture, the complexity of  a {$|\psi(t_L,t_R)\rangle$} is dual to the action in the Wheeler-DeWitt (WDW) patch associated with $t_L$ and $t_R$, i.e.
\begin{equation}\label{CA}
  \mathcal{C}_A=\frac{I_{\text{WDW}}}{\pi\hbar}.
\end{equation}
The WDW patch associated with $t_L$ and $t_R$ is the set of all space-like surfaces connecting $t_L$ and $t_R$ with the null sheets coming from $t_L$ and $t_R$. More precisely it is the domain of dependence of any space-like surface connecting $t_L$ and $t_R$. This conjecture has some advantages compared with the CV conjecture. For example, it has no free parameter and can satisfy Lloyd's complexity growth bound in some cases \cite{Lloyd2000,Cai:2016xho,Yang:2016awy,Cai:2017sjv}. However, the CA conjecture has its own obstacle in computing the action: it involves null boundaries and joint terms. This problem has been overcome by carefully analyzing the boundary term in null boundary \cite{Parattu:2015gga,Lehner:2016vdi}.

Compared with the incessant progresses from gravity duality (see Refs.~\cite{Ghodrati:2017roz,Hashimoto:2017fga,Alishahiha:2015rta} for some recent progresses by gravity duality), it seems that we meet stiff obstacle in giving a well definition about the complexity in pure field theory framework. This difficulty partly comes from its original idea. Roughly speaking, the complexity characterizes how difficult it is to obtain a particular quantum state from an appointed reference state.  In a discrete system, such as a quantum circuit, it's the minimal number of required gates to convert the reference state into a particular state \cite{2008arXiv0804.3401W,2014arXiv1401.3916G,2012RPPh...75b2001O}. This idea cannot directly be applied into the continuous system.

In order to find a kind of generalization about gate complexity, Nielsen and collaborators~\cite{Nielsen1133,Nielsen:2006:GAQ:2011686.2011688,Dowling:2008:GQC:2016985.2016986} constructed a continuum approximation to gate complexity which involved a new kind of ``complexity geometry." In Nielsen's works, the complexity is geometrized by Finsler geometry~(some introduction about Finsler geometry can be found in Refs.~\cite{038798948X,9810245319}). The Finsler geometry depends on the choice of Finsler structure. Different choices on Finsler structure may lead to different results. At current, it seems that there is no any method to determine the Finsler structure uniquely. Recently, Ref.~\cite{Jefferson:2017sdb} used some different Finsler structures to compute the complexity for some systems and showed some similarities compared with the CV and CA conjectures. However, as the systems they checked are very different from TFD states, their results and holographic results have still some differences. Therefor, if we really want to compare the complexity based on a field theory framework and holography, we need  apply the field theory framework into TFD states. This is what this paper will focus on.

On the other hand, though some positive results have been obtained from the CV and CA conjectures, the understanding on them is still at the very preliminary stage. It is still not clear that if CV and CA conjectures are completely correct. As the complexity depends on the reference state, it is an important and fundamental question to clarify the reference state. However, CV and CA conjectures themselves do not tell us what the reference state is. It is possible that the disappearance  of reference state in these two conjectures is because they are not the complete versions and some modifications may be needed. These questions obviously cannot obtain the answers only by CV and CA conjectures themselves. A well definition and full study about complexity based on pure field theory are needed. This is also one motivation of this paper.

There is also a very surprising coincidence in holography complexity and the holographic conjecture about fidelity susceptibility~\cite{MIyaji:2015mia,Alishahiha:2017cuk}. The fidelity is also a very important conception in quantum information theory, which measures similarity of two states (a brief introduction about the fidelity and fidelity susceptibility will be shown in subsection~\ref{Fideli2C}. For more details, one can refer to Ref.~\cite{FIDELITY}). The fidelity susceptibility and complexity, in principle, are two different conceptions. Ref.~\cite{MIyaji:2015mia} gives a holographic description and says that its gravity dual is approximately given by the maximal volume of time slice in an AdS spacetime, which shares the same object with the holographic complexity in CV conjecture. This coincidence seems to imply that, at least for TFD states, the complexity and fidelity susceptibility have some deep connection and may be equivalent to each other.  To answer this question and clarify why such coincidence can happen, we also need a well defined quantum field theory proposal for complexity.

This paper will study the complexity of states in quantum field theory by introduce a Finsler structure based on the ladder operators (the generalization of creation and annihilation operators). In the Sec.~\ref{Comgeo}, some basic properties of complexity will be proposed and a method to construct the Finsler structure will be presented. Then this method will be first applied into some simple examples in Sec.~\ref{sim-exmp}. Especially, the complexity of coherent states and entanglement thermofield states are computed as examples to show how to use this method. These examples will also clarify some differences between complexity and other conceptions such as complexity of formation and entanglement entropy. In Sec.~\ref{Com-TFD}, this method is applied to compute complexity between thermofield double states.  The results show that the complexity density between a thermofield double state and corresponding zero temperature ground state is finite.  In addition,  it is found that complexity $\mathcal{C}$ and temperature $T$ in $d$-dimension free conformal field theory shows the behavior of $\mathcal{C}\propto T^{d-1}$, which is just the the renormalized complexity predicted by CA and CV conjectures~\cite{Kim:2017lrw}. Especially, an explicit proof will be given to show fidelity susceptibility of a TFD state is equivalent to the complexity between it and corresponding vacuum state, which gives an explanation on why they may share the same object in holography. In Sec.~\ref{whyfinite}, some physical discussions will be found to explain why the complexity between a TFD state and its corresponding zero temperature vacuum state should be finite, and some comments on understanding about CV and CA conjectures will also be given. A short summary and some outlooks will be given in Sec.~\ref{summ}.

\section{Complexity geometry}\label{Comgeo}
\subsection{Geometrization of complexity}
In the works of Refs.~\cite{Nielsen1133,Nielsen:2006:GAQ:2011686.2011688,Dowling:2008:GQC:2016985.2016986}, instead of to directly construct the complexity between states, they first defined the complexity of operators\footnote{The other idea was proposed by Ref.~\cite{Chapman:2017rqy}, which defined a line element in Hilbert space by Fubini-Study metric~\cite{bengtsson2006geometry}.}. Here I will follow this idea and propose some basic properties of complexity. 

\begin{figure}
  \centering
  \includegraphics[width=.4\textwidth]{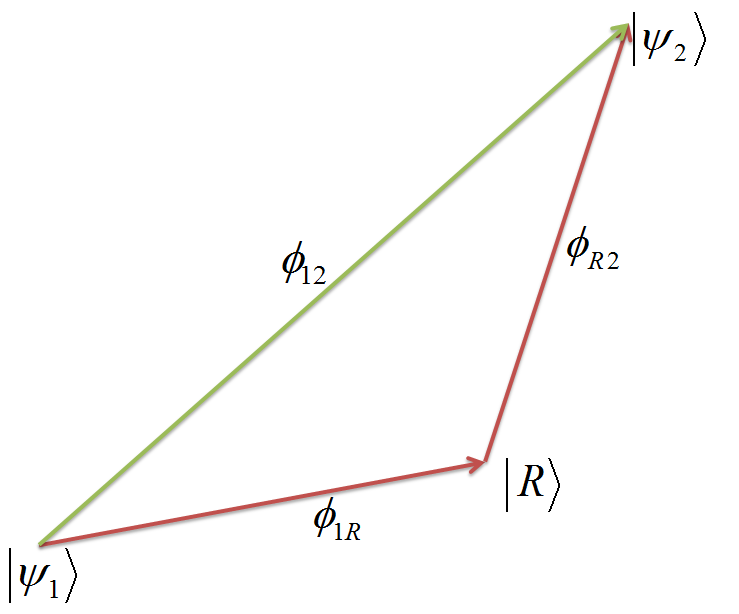}
  \caption{The schematic explanations about the triangle inequality. $\phi_{1R}$ is one quantum circuit of minimal gates to realize $|\psi_1\rangle\rightarrow|R\rangle$, $\phi_{R2}$ is one quantum circuit of minimal gates to realize $|R\rangle\rightarrow|\psi_2\rangle$, and $\phi_{12}$ is one quantum circuit of minimal gates to realize $|\psi_1\rangle\rightarrow|\psi_2\rangle$. As the combination $\phi_{R2}\circ\phi_{1R}$ is a possible quantum circuit to realize $|\psi_1\rangle\rightarrow|\psi_2\rangle$ with the gates number $\mathcal{C}(|\psi_2\rangle,|R\rangle)+\mathcal{C}(|R\rangle,|\psi_1\rangle)$ which should be larger than or equal to the gates number of $\phi_{12}$, we see that $\mathcal{C}(|\psi_2\rangle,|R\rangle)+\mathcal{C}(|R\rangle,|\psi_1\rangle)\geq\mathcal{C}(|\psi_2\rangle,|\psi_1\rangle)$. }\label{ineqs1}
\end{figure}

For the case that all the admitted operators form a continuous manifold $\mathcal{U}$, they introduced of a Finsler structure $\mathcal{F}$, which is a non-negative function defined on its tangent bundle $T\mathcal{U}$. For any piecewise $C^1$ curve $\hat{c}:[0,1]\mapsto\mathcal{U}$ which satisfies $\hat{c}(0)=I$ and $\hat{c}(1)\in\mathcal{U}$, one can define its length $L[\hat{c}]$ such that,
\begin{equation}\label{leghtL}
  L[\hat{c}]:=\int_0^1\td t\mathcal{F}\left[\hat{c}(t),\hat{T}(t)\right]\,.
\end{equation}
Here $\hat{T}(t)$ is the tangent of the curve and satisfies that $\frac{\td}{\td t}\hat{c}(t)=\hat{T}(t)\hat{c}(t)$. Then Nielsen defined the complexity of $\hat{U}$ by\footnote{In the original definition of Nielsen's, the $\lambda$ has to be 1. However, it can be relaxed that $\lambda$ is any nonzero complex number in this paper, as we here only consider the operators which are acted on quantum states. For any quantum state $|\psi\rangle$, $\hat{U}|\psi\rangle$ and $\lambda \hat{U}|\psi\rangle$ describe the same state. },
\begin{equation}\label{defCU1}
  \mathcal{C}(\hat{U})=\min\left\{\left.L[\hat{c}]~\right|~\forall \hat{c}:[0,1]\mapsto\mathcal{U}, \exists\lambda\neq0, s.t., \hat{c}(0)=\hat{I}, \hat{c}(1)=\lambda \hat{U}\right\}\,.
\end{equation}
Here $\hat{I}$ is the the identity of $\mathcal{U}$. This definition leads to following two properties:  $\forall \hat{U},\hat{U}_1,\hat{U}_2\in\mathcal{U}$\\
(1a) $\mathcal{C}(\hat{U})=0\Leftrightarrow \exists\lambda\neq0,.s.t., \hat{U}=\lambda \hat{I}$;\\
(2a) Subadditivity: $\mathcal{C}(\hat{U}_1)+\mathcal{C}(\hat{U}_2)\geq\mathcal{C}(\hat{U}_1\hat{U}_2)$ if $\hat{U}_1\hat{U}_2\in\mathcal{U}$.\\
The complexity of states then can be defined based on the complexity of operator. For any two states $|\psi_1\rangle$ and $|\psi_2\rangle$, the complexity from $|\psi_1\rangle$ to $|\psi_2\rangle$ can be defined by following way,
\begin{equation}\label{defcompstates}
  \mathcal{C}(|\psi_2\rangle,|\psi_1\rangle)=\min\{\mathcal{C}(\hat{U}_i)~|~\forall \hat{U}_i\in\mathcal{U}, s.t., |\psi_2\rangle\sim \hat{U}_i|\psi_1\rangle\}\,.
\end{equation}
Here notation ``$\sim$'' means that the two sides can differ from each other up to any nonzero complex number. This definition leads to following two properties,\\
(1b) $\mathcal{C}(|\psi_2\rangle,|\psi_1\rangle)=0$ if and only if $|\psi_2\rangle\sim|\psi_1\rangle$;\\
(2b) Triangle inequality: $\mathcal{C}(|\psi_2\rangle,|R\rangle)+\mathcal{C}(|R\rangle,|\psi_1\rangle)\geq\mathcal{C}(|\psi_2\rangle,|\psi_1\rangle)$ for any state $|R\rangle$.\\
One understanding and proof for (2b) can be found in Fig.~\ref{ineqs1}. The complexity does not have reversibility in general, i.e., $\mathcal{C}(|\psi_2\rangle,|\psi_1\rangle)\neq\mathcal{C}(|\psi_1\rangle,|\psi_2\rangle)$.

There is also a useful conception named ``complexity of formation'' proposed by Ref.~\cite{Chapman:2016hwi}. It describes what is the additional complexity arising in preparing state $|\psi_1\rangle$ compared with $|\psi_2\rangle$ from a reference state $|R\rangle$. This can be defined by $\Delta\mathcal{C}_R(|\psi_2\rangle,|\psi_1\rangle)$ in following way,
\begin{equation}\label{formC}
  \Delta\mathcal{C}_R(|\psi_2\rangle,|\psi_1\rangle):=\mathcal{C}(|\psi_2\rangle,|R\rangle)-\mathcal{C}(|\psi_1\rangle,|R\rangle)\,.
\end{equation}
In general, the complexity of formation between the two states depends on the choice of reference state. Especially, when we choose that the reference state $|R\rangle$ is the state $|\psi_1\rangle$, then ``complexity of formation'' just gives the complexity from $|\psi_1\rangle$ to $|\psi_2\rangle$. In general cases, the property (2b) shows that,
\begin{equation}\label{formC2}
  \mathcal{C}(|\psi_2\rangle,|\psi_1\rangle)\geq\Delta\mathcal{C}_R(|\psi_1\rangle,|\psi_2\rangle)~~~\text{for}~\forall|R\rangle\,.
\end{equation}
The inequality~\eqref{formC2} can not be strengthened into the $\mathcal{C}(|\psi_2\rangle,|\psi_1\rangle)\geq|\Delta\mathcal{C}_R(|\psi_1\rangle,|\psi_2\rangle)|$ in general as the complexity may not have reversibility. In general, the complexity of formation and complexity from one to the other are different. However, it will be shown in the subsection~\ref{ETS1} that they can be equivalent in some special cases.

\begin{figure}
  \centering
  \includegraphics[width=.4\textwidth]{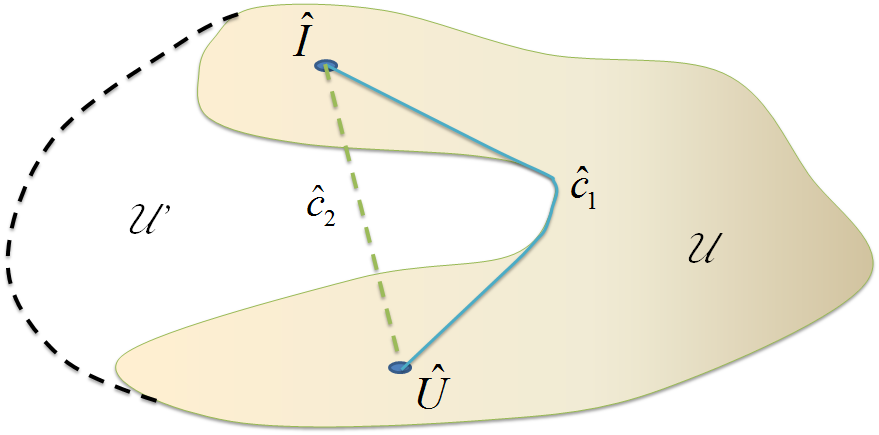}
  \caption{The schematic example that the value of $\mathcal{C}(\hat{U})$ depends on  the choice of $\mathcal{U}$. The operators set $\mathcal{U}$ is a subset of a larger operators set $\mathcal{U}'$. The curve length is just given by Euclidean metric. For the operators set $\mathcal{U}$, the shortest curve from $\hat{I}$ to $\hat{U}$ is given by $\hat{c}_1$. However, if one extend the the operators set $\mathcal{U}$ to $\mathcal{U}'$, the shortest curve from $\hat{I}$ to $\hat{U}$ becomes $\hat{c}_2$. This shows that the complexity of $\hat{U}$ depends on the choice of operators set.}\label{Udeps}
\end{figure}

It needs to emphasis three properties based on the definitions~\eqref{defCU1} and \eqref{defcompstates}. The first one is that we must first state what is the admitted operators set $\mathcal{U}$ as the value of $\mathcal{C}(\hat{U})$ defined by \eqref{defCU1} depends on the choice of $\mathcal{U}$ (see Fig.~\ref{Udeps} as an example).  The second one is that in general we can not say $\mathcal{C}(|\psi_2\rangle,|\psi_1\rangle)=\mathcal{C}(|\psi_1\rangle,|\psi_2\rangle)$. In fact, this point emerges from the physical intuition very naturally, as we can feel that the costs for many processes and their inverses are different. The third point is that there may be many different operators $\hat{U}_i$ which can satisfy the relationship $|\psi_2\rangle\sim \hat{U}_i|\psi_1\rangle$. We should compare the complexities of all these operators by using Eq.~\eqref{defCU1} and find the minimal value of them to determine the complexity from $|\psi_1\rangle$ to $|\psi_2\rangle$.

Refs.~\cite{Nielsen1133,Nielsen:2006:GAQ:2011686.2011688,Dowling:2008:GQC:2016985.2016986} studied the complexity in special operators set $\mathcal{U}_E$, where all the curves can be generated by some ``time-dependent Hamiltonians'' $\hat{T}(t)$ such that,
\begin{equation}\label{curveH1}
  \hat{c}(s)=\overleftarrow{\mathcal{P}}\exp\int_0^s\hat{T}(t)\td t\,
\end{equation}
and the $\overleftarrow{\mathcal{P}}$ indicates a time ordering such that the Hamiltonian at earlier times is applied to the state first. Any ``time-dependent Hamiltonian'' $\hat{T}(t)$ should be expanded in the basis $E=\{\hat{M}^1, \hat{M}^2, \cdots\}$ such that,
\begin{equation}\label{expandH}
  \hat{T}(t)=Y_i(t)\hat{M}^i\,.
\end{equation}
The basis $E$, which can be treated as the generators of operators set  $\mathcal{U}_E$ and plays the role of universal set (minimal complete gates set) in the quantum gates, determines the operators set $\mathcal{U}_E$.\footnote{Here it does not need that  $\mathcal{U}_E$ forms any group} Then the Finsler structure $\mathcal{F}$ in Eq.~\eqref{leghtL} can be given by a basis-dependent function $F$ in this way,
\begin{equation}\label{Finsler1}
  \mathcal{F}\left[\hat{c}(t),\hat{T}(t)\right]=F[\hat{c}(t);Y_I(t)]\,.
\end{equation}
Here the Finsler structure and its function form expressed in a basis, i.e., $\mathcal{F}$ and $F$,  should be distinguished. The reason will be explained later on.
Ref.~\cite{Nielsen:2006:GAQ:2011686.2011688} studed different types of Finsler structure $\mathcal{F}$ and compared their advantages and disadvantages. We will return to it in the Sec.~\ref{F-struc}.

In general, one can choose different bases and obtain different admitted  operators sets. For example, one can choose a larger basis $E'\supset E$ so that we can generate a larger operators set $\mathcal{U}_{E'}$ and some new curves. Then we see that $\mathcal{C}(U)|_{\mathcal{U}_E}\geq\mathcal{C}(U)|_{\mathcal{U}_{E'}}$.  Let's assume $E'=\{\hat{M}'^1,\hat{M}'^2,\cdots\}$ and $E=\{\hat{M}^1,\hat{M}^2,\cdots\}$. For the special case $E$ and $E'$ can be associated by a linear transformation, i.e., there is a matrix such that,\footnote{In this paper, the Einstein summation rule has been used for upper and lower indexes, i.e., it needs to make a summation if the same index notation appears in the upper and lower indexes.}
\begin{equation}\label{E1toE2}
  \hat{M}'^I={A^I}_J\hat{M}^J\,.
\end{equation}
Here ${A^I}_J$ may be not an investible matrix. Then we can see that $\mathcal{U}_{E'}\subseteq\mathcal{U}_{E}$. If the tangent $\hat{T}(t)$ is also tangent to $\mathcal{U}_{E'}$, then it can be expanded by the new basis $E'$ such as $\hat{T}(t)=Y'_I\hat{M}'^I$. The coefficients $Y'_I$ and $Y_I$ will has following relationship,
\begin{equation}\label{tranforY}
  Y_I={{A}^J}_IY'_J\,.
\end{equation}
Assume $\mathcal{F}'$ to be the Finsler structure in $\mathcal{U}_{E'}$ and $F'$ is its function form defined in the basis $E'$. As $\mathcal{U}_{E'}\subseteq\mathcal{U}_{E}$, we can compute the complexity for any operator $\hat{U}\in\mathcal{U}_{E'}$ by two different Finsler structures $\mathcal{F}'$ and $\mathcal{F}$. If we require two Finsler structures can give the same length of the curve generated by $\hat{T}(t)$, then their function forms in the odd and new bases should satisfies following condition,
\begin{equation}\label{twoFs}
  F'[\hat{c}(t);Y'_J]=F[\hat{c}(t);{A^I}_JY'_I], ~~~\forall~Y'_I\,.
\end{equation}
This gives the transformation rule for function form of Finsler structure under the basis transformation~\eqref{E1toE2}. We see that, for a given Finsler structure $\mathcal{F}$, its function form depends on the choice of basis. Because of this reason,  we have to distinguish Finsler structure $\mathcal{F}$ and its function form $F$. In the appendix~\ref{transF1}, I will give an example about how to use this transformation rule to obtain the function forms of Finsler structure in new bases.

To give the definition \eqref{defCU1} a well meaning, we have to first appoint the basis (i.e., the generators set) $E$ and Finsler structure function form $F$ corresponding to this basis. Then the pair $(E,F)$ determines the complexity of any operator in set $\mathcal{U}_E$ and the complexity between two states in set $\mathcal{S}_E$, where $\mathcal{S}_E$ is the states set in which any two states can be transformed by the operators in $\mathcal{U}_E$. In general, the states set $\mathcal{S}_E$ is not the whole Hilbert space $\mathcal{H}$, as there may be two states in  $\mathcal{H}$ which cannot be transformed from one to the other by the operators in $\mathcal{U}_E$.

Let's make a short argument on why the Finsler geometry is a natural generalization of gates complexity in the continuous system.  Finding the complexity of an operator $\hat{U}$ in both the classical and quantum circuits can be concluded in following steps. Firstly, one has to choose a universal gates set (the fundamental components in constructing the circuit) $E=\{\hat{e}_1,\hat{e}_2,\cdots\}$. By repeating to use these components, one can find ways to construct some quantum/calssical circuits to realize the operator $\hat{U}$. There may be many different ways to construct different circuits to realize the same operator $\hat{U}$. To determine which design is optimal, one has to appoint the ``cost'' $F_i>0$ for every element $\hat{e}_i\in E$. Then the total cost of the design is the summation of the cost of every element, i.e., $\sum F_i$. If we appoint that $F_i$ is 1 for all the components, then the total cost is just the gates number in the circuit. In general, the optimal design is the one which can make $\sum F_i$ minimal, and this minimal value is the complexity of operator $\hat{U}$. The pair $(E,F)$ in continuous case is almost the continuous version of pair $(E,F_i)$ in claasical/quantum circuit.




\subsection{Finsler structure and generators}\label{F-struc}
In Ref.~\cite{Jefferson:2017sdb} (and also in the previous works of Nielsen's), the Finsler structures were constructed by paying more attention to the group structure itself. In this paper, I will try consider the problem from how to construct the states in Hilbert space. One will see later that this viewpoint will be very suitable to study the complexity between quantum states, especially for the TFD states.

Let's consider a bosonic Hilbert space $\mathcal{H}$. For convenience, let's assume the system has discrete momentum.
To describe a state in $\mathcal{H}$, we need to choose a representation, i.e., a series of basic vectors. One common choice in free theory is the particle number operator,
\begin{equation}\label{particleN}
  \hat{N}:=\hat{a}^\dagger\hat{a}\,.
\end{equation}
Here operator $\hat{a}^\dagger$ and  $\hat{a}$ are the creation and annihilation operators. In most cases, there are some different creation and annihilation operators, which are commutative to each others. To distinguish such different particles created by different creation operators, we can add some indexes. For example, let's consider the case that particles can carry different momentum. Then the particle number density operators corresponding to momentum $\vec{k}_i$ is,
\begin{equation}\label{particle2}
  \hat{N}_{\vec{k}_i}:=\hat{a}^\dagger_{\vec{k}_i}\hat{a}_{\vec{k}_i}\,.
\end{equation}
Here operators $\hat{a}^\dagger_{\vec{k}}$ and  $\hat{a}_{\vec{k}}$ are the creation and annihilation operators, which can add a particle of momentum $\vec{k}$ or annihilate a particle of momentum $\vec{k}$. Let's use the notation,
\begin{equation}\label{basics1}
  \prod_{i=0}|n_i,\vec{k}_i\rangle:=|n_0,\vec{k}_0\rangle|n_1,\vec{k}_1\rangle|n_2,\vec{k}_2\rangle\cdots
\end{equation}
to stand for the state that there are $n_0$ particles of momentum $\vec{k}_0$,  $n_1$ particles of momentum $\vec{k}_1$, $n_2$ particles of momentum $\vec{k}_2$,$\cdots$. The state in Eq.~\eqref{basics1} is the common eigenvector of all the particle number density operators and can form a complete basis in Hilbert space $\mathcal{H}$. Any state in the Hilbert space $\mathcal{H}$ can be presented as,
\begin{equation}\label{stateN1}
  |\psi\rangle=\sum_{n_0,n_1,\cdots=0}^\infty c_{n_0n_1\cdots}\prod_{i=0}|n_i,\vec{k}_i\rangle=\sum_{n_0,n_1,\cdots=0}^\infty c_{n_0n_1\cdots}|n_0,\vec{k}_0\rangle|n_1,\vec{k}_1\rangle\cdots\,.
\end{equation}
This is the occupation number representation of a state in Hilbert space, which is one basic representation in canonical quantization in quantum field theory and second quantization in quantum many-body systems. The coefficient matrix $c_{n_0n_1n_2\cdots}$  can be reorganized into the matrix product state presentation~\cite{2006quant.ph..8197P} or multi-scale entanglement renormalization
ansatz (MERA) \cite{PhysRevLett.115.200401}.

The physical meanings of operators $\hat{a}_{\vec{k}}$ and $\hat{a}^\dagger_{\vec{k}}$ are very clear: they stand for deleting or adding one particle of momentum $\vec{k}$ in the system. If the bosonic field is the fundamental field rather then an effective field, then  particles are indivisible. It is very naturally to regard that the fundamental operators are adding and deleting one particle. This advises us to choose following generators set,
\begin{equation}\label{QFTbasis}
  E^{0}:=\bigcup_{i}\{\hat{a}^\dagger_{\vec{k}_i}, \hat{a}_{\vec{k}_i}, \hat{\mathbb{I}}\}
\end{equation}
%
Here $\hat{\mathbb{I}}:=[\hat{a}_{\vec{k}_i}, \hat{a}^\dagger_{\vec{k}_j}]\delta_{ij}$ is the center of $E^{0}$ and satisfies $\hat{\mathbb{I}}\hat{e}=\hat{e}$ for $\forall\hat{e}\in E^{0}$. The generator set $E^{0}$ forms an infinite dimensional Heisenberg–Weyl Lie algebra. In general, this basis is not big enough for the physical interesting questions, so let's extend the the basis in this way,
\begin{equation}\label{extendbasis}
  E:=\bigcup_{n=1}^\infty(E^0)^n, ~~~\text{with}~(E^0)^n:=\{\hat{M}^{i_1i_2\cdots i_n}=:\hat{e}_1\hat{e}_2\cdots \hat{e}_n:|\forall \hat{e}_1,\hat{e}_2,\cdots,\hat{e}_n\in E^0\}\,.
\end{equation}
Here the ``:~:'' means that annihilation operators will always appear at the right of corresponding creation operator, e.g., $:\hat{a}_{\vec{k}_i}\hat{a}^\dagger_{\vec{k}_i}:=\hat{a}^\dagger_{\vec{k}_i}\hat{a}_{\vec{k}_i}$. In the definition~\eqref{extendbasis}, $\hat{e}_1,\hat{e}_2,\cdots,\hat{e}_n$ do not need to be different from each others. Such extended basis in fact is nothing but the universal enveloping algebra of Heisenberg–Weyl Lie algebra.\footnote{In mathematics, the universal enveloping algebra of set $E$ defined by Eq.~\eqref{extendbasis} can be induced by Heisenberg–Weyl Lie algebra, i.e., the basic commutative relationship $[\hat{a}_{\vec{k}}, \hat{a}^\dagger_{\vec{k}'}]=\delta_{\vec{k}\vec{k}'}\hat{\mathbb{I}}$. Thus it also forms a Lie algebra and the corresponding operators set is also a Lie group. }

For free field theory, the particle number is conversed and very convenient to characterize the quantum states. However, in interacted field theories, the total Hamiltonian  $H$ and the particle number density operator $\hat{N}_{\vec{k}_i}$ defined in Eq.~\eqref{particle2} are not commutative, so the the particle number is not conversed. In this case, we can use ladder operators $\{\hat{l}_{\vec{k}_i},\hat{l}^\dagger_{\vec{k}_i}\}$ (the generalization creation/anihinlation operators) to replace the creation/anihinlation operators $\{\hat{a}^\dagger_{\vec{k}_i}, \hat{a}_{\vec{k}_i}\}$. For arbitrary field theory, let $\hat{H}$ be its Hamilton which has discreted eigenvalues $E_n$ and is commutative to momentum operator. In the appendix~\ref{app1}, I will prove that there is a unique operators set $\{\hat{l}_{\vec{k}_i},\hat{l}^\dagger_{\vec{k}_i}\}$ which can satisfies,
\begin{equation}\label{ladders}
  [\hat{l}_{\vec{k}'},\hat{l}_{\vec{k}}]=[\hat{l}^\dagger_{\vec{k}'},\hat{l}^\dagger_{\vec{k}}]=0,~~[\hat{l}_{\vec{k}'},\hat{l}^\dagger_{\vec{k}}]=\hat{\mathbb{I}}\delta_{\vec{k}',{\vec{k}}}, ~~\hat{l}_{\vec{k}}|E_n,\vec{k}\rangle=\alpha_{n,\vec{k}}|E_{n-1},\vec{k}\rangle\,
\end{equation}
with $\alpha_{n,\vec{k}}>0$ if $n>0$ and $\alpha_{0,\vec{k}}=0$. The vacuum state corresponding to this Hamilton then is $|0\rangle:=\prod_{i=0}|E_{0},\vec{k}_i\rangle$.
The operator $\hat{l}^\dagger_{\vec{k}_i}$ can change the energy at momentum $\vec{k}$ from $E_n$ to $E_{n+1}$. By applying the  $\hat{l}^\dagger_{\vec{k}_i}$, we can create the any state $\prod_{\vec{k}_i}|E_{n_i},\vec{k}_i\rangle$ from vacuum state and the set $\cup_{\{n_i\}}\{\prod_{i=0}|E_{n_i},\vec{k}_i\rangle\}$ forms a complete basis in Hilbert space $\mathcal{H}$. The operators $\{\hat{l}_{\vec{k}_i},\hat{l}^\dagger_{\vec{k}_i}\}$ are called the ladder operators corresponding to Hamilton $\hat{H}$. $\hat{l}_{\vec{k}_i}$ is the lowering operator (the generalized annihilation operator) and $\hat{l}^\dagger_{\vec{k}_i}$ is the raising operator (the generalized creation operator).  One can also find that the generalized particle number density operator $$\hat{N}'_{\vec{k}}:=\hat{l}^\dagger_{\vec{k}}\hat{l}_{\vec{k}}$$
is commutative to the Hamilton $\hat{H}$. Thus, in general system, we can use the ladder operators to replace the creation and annihilation operators and define,
\begin{equation}\label{QFTbasis2}
  E^{0}:=\bigcup_{i}\{\hat{l}^\dagger_{\vec{k}_i}, \hat{l}_{\vec{k}_i}, \hat{\mathbb{I}}\}\,.
\end{equation}
It still forms an infinite dimensional Heisenberg–Weyl Lie algebra. If the system is given by a free theory, then the ladder operators are just the creation and annihilation operators.  For convenience, we will still use the notation $\{\hat{a}^\dagger_{\vec{k}_i}, \hat{a}_{\vec{k}_i}\}$ to stand for ladder operators and $|n\rangle$ to stand for the energy eigenstate $|E_n\rangle$ in interacted systems. The readers can keep in mind that $\{\hat{a}^\dagger_{\vec{k}_i}, \hat{a}_{\vec{k}_i}\}$ stands for the ladder operators corresponding to total Hamilton  when we discuss the interacted theory.

After we have prepared the basis already, then the tangent vector $\hat{T}(t)$ can be decomposed in this way,
\begin{equation}\label{decompH1}
  \hat{T}(t)=T_0(t)\hat{\mathbb{I}}+\sum_i Y_i(t)\hat{M}^i+\sum_{ij}Y_{ij}(t)\hat{M}^{ij}+\cdots+\sum_{i_1i_2\cdots i_n}Y_{i_1i_2\cdots i_n}(t)\hat{M}^{i_1i_2\cdots i_n}+\cdots\,.
\end{equation}
Here $T_0(t), Y_i, Y_{ij}(t),\cdots$ are complex numbers, $\hat{M}^i, \hat{M}^{ij},\cdots$ are the generators given by Eq.~\eqref{extendbasis} and not the center $\hat{\mathbb{I}}$. In addition, we can require that the tangent operator $\hat{T}(t)$ should be anti-Hermit, i.e., $\hat{T}(t)^\dagger=-\hat{T}(t)$, so that the operator generated by it is unitary. This requirement is natural in physics. However, as this paper is going to explain the basic idea about how to construct the complexity in quantum field theory, we will not add this requirement.

Now we have to appoint the function form of Finsler structure in this basis. Four different types of $F$ have been studied in Ref.~\cite{Nielsen:2006:GAQ:2011686.2011688} and also been checked in recent paper~\cite{Jefferson:2017sdb}, which are,
\begin{equation}\label{fourFs}
  \begin{split}
  F_1&=\sum_I\parallel Y_I\parallel ,~~~F_{p}=p^I\parallel Y_I\parallel \\
  F_2&=\sqrt{\sum_I|Y_I|^2},~~~F_{q}=\sqrt{q^I|Y_I|^2}\,.
  \end{split}
\end{equation}
The summation includes all the indexes of $Y$ in Eq.~\eqref{decompH1}. In the two Finsler structures on the right side, $p^I$ and $q^I$ are penalty factors which can be chosen to favour certain ones in the fundamental generators/gates over others, i.e., to give a higher cost to certain classes
of gates. For real numbers, the notation $\parallel \cdot\parallel $ is defined as $\parallel Y^I\parallel :=|Y^I|$, i.e., the usual absolute value of real number. As what have been compared in quantum circuits formed by spin chain in Ref.~\cite{Nielsen:2006:GAQ:2011686.2011688}, $F_1$ is the best motivated of all the four local Finsler structures. One physical interpretation for such preference is as follow.  Suppose $\hat{U}$ to be generated by applying sequentially the discrete fundamental operators (logic gates) which are generated by ${\hat{\sigma}_1}, {\hat{\sigma}_2},\cdots$ at the time $t_1, t_2, \cdots$. Then we can use $\delta$-function to write these discrete operators into a generator as following form,
\begin{equation}\label{deltaH}
  \hat{T}(t)=\delta(t-t_1)\hat{\sigma}_1+\delta(t-t_2)\hat{\sigma}_2+\cdots=\sum_{n=0}\delta(t-t_n)\hat{\sigma}_n.
\end{equation}
$F_1$ leads that the length defined by Eq.~\eqref{leghtL} for this ``curve''  is just the total number of fundamental operators, so the curve of minimal length just corresponds to the design of minimal required gates. In this sense, $F_1$ is the most natural generalization of the gate complexity for continuous system. $F_p$ is a modified version of $F_1$ in which we introduce a penalty for some generators. But functions $F_1$ and $F_p$ cannot give Finsler structures in strict sense. However, Ref.~\cite{Nielsen:2006:GAQ:2011686.2011688} shows that this can be overcome by treating them as the limit of some continuous function. Thus, this subtlety will not be important in physics.

When the coefficients $Y^I$ in Eq.~\eqref{decompH1} are complex number, the notation $\parallel \cdot\parallel $ is a little ambiguous as  $Y^I=\rho^I e^{i\theta^I}$ in fact stand for two numbers $(\rho^I,\theta^I)$ rather than one number. Naively thinking, we should use $\parallel Y^I\parallel =\rho^I$. However, this naive idea in fact is against the original intention of $F_1$, as the ``rotation'' caused by $\theta^I$ is not counted into the complexity. One way to generalize $F_1$ for complex number $Y^I=\rho^I e^{i\theta^I}$ is,
\begin{equation}\label{twoF1}
  \parallel Y^I\parallel :=\rho^I(|\cos\theta^I|+|\sin\theta^I|\cdot|\theta^I|)\,.
\end{equation}
The simple physical meaning for this generalization is as follows.  As $Y^IM_I=\rho^I\cos\theta^I M_I+\rho^I\sin\theta^I\cdot (i\cdot M_I)$, we can treat $M_I$ and $(i\cdot M_I)$ as different generators and give additional wight $|\theta|$ to $(i\cdot M_I)$ by the consideration that it makes the $Y^I$ to rotate $|\theta|$ angular. For a given $Y^I(t)$, the value of $\theta(t)$ is not unique. To avoid this ambiguous, we can require that $\theta\in[-\pi,\pi)$ for constant $\theta$. If $\theta(t)$ is not a constant, then we require $\theta(0)\in[-\pi,\pi)$ and $\theta(t)$ is continuous when $t\in[0,1]$. As $ \parallel Y^I\parallel$ is the even function, we can take $\theta(0)=\arccos(\text{Re}Y^I/\rho)$.

In this paper, we will use $F_p$ Finsler structure function form. For the tangent vector $\hat{T}(t)$ shown in Eq.~\eqref{decompH1}, it is naturally to introduce function $F$ in the basis $E$ in this way,
\begin{equation}\label{defFp10}
  F=\ell\left[p\parallel T_0(t)\parallel+\sum_i \parallel Y_i(t)\parallel +2\sum_{ij}\parallel Y_{ij}(t)\parallel +\cdots+n\sum_{i_1i_2\cdots i_n}\parallel Y_{i_1i_2\cdots i_n}(t)\parallel +\cdots\right]\,.
\end{equation}
Here $\ell$ is a free parameters and positive. One can prove that in order to match the requirement that $\mathcal{C}(\hat{I})=0$, we have to set that $$p=0.$$
It is every naturally to choose the weight factors for other coefficients as Eq.~\eqref{defFp10}, as the generator $\hat{M}^i$ only contains the operators which can create or annihilate one particle, generator $\hat{M}^{i_1i_2\cdots i_n}$ only contains the operators which can create or annihilate $n$ particles. The decomposition Eq.~\eqref{decompH1} can be generalized into the continuous cases.  The discrete annihilation operator $\hat{a}_{\vec{k}_i}$ and its continuous form have the relation,
\begin{equation}\label{adtoac}
  \sqrt{\frac{\text{Vol}}{(2\pi)^{d-1}}}\hat{a}_{\vec{k}_i}\rightarrow\hat{a}(\vec{k})\,.
\end{equation}
as well as the relationship between the summation and integration,
\begin{equation}\label{sumint}
  \sum_{\vec{k}_i}\rightarrow\frac{\text{Vol}}{(2\pi)^{d-1}}\int\td^{d-1}k\,.
\end{equation}
Here Vol stands for the volume of the space where the field can distribute and $d$ is the spatial dimensions of corresponding quantum field theory. The function form of Finsler structure then becomes,
\begin{equation}\label{defFp1}
\begin{split}
  \ell^{-1}F&=\left[\frac{\text{Vol}}{(2\pi)^{d-1}}\right]\int\td^{d-1}k\parallel Y_{\vec{k}}(t)\parallel +2\left[\frac{\text{Vol}}{(2\pi)^{d-1}}\right]^2\iint\td^{d-1}k_1\td^{d-1}k_2\parallel Y_{\vec{k}_1\vec{k}_2}(t)\parallel \\
  &+\cdots+n\left[\frac{\text{Vol}}{(2\pi)^{d-1}}\right]^{n}\left(\prod_{i=1}^{n}\int\td^{d-1}k_i\right)\parallel Y_{\vec{k}_1\cdots\vec{k}_n}(t)\parallel +\cdots
  \end{split}
\end{equation}

After we have prepared the pair $(E, F)$, then we can compute the complexity of any operators in $\mathcal{U}_E$ and complexity between the states in $\mathcal{S}_E$. It is not clear that if the states set $\mathcal{S}_E$ can contain the all the states in the whole Hilbert space $\mathcal{H}$. However, it will be show that the TFD states, which are the main targets in the holographic duality, are contained in the states set $\mathcal{S}_E$. In addition, the states studied by Ref.~\cite{Chapman:2017rqy} are also contained in $\mathcal{S}_E$. For convenience, the following sections of this paper will take $\ell=1$.

\section{Complexity in some simple examples}\label{sim-exmp}
\subsection{Complexity  between coherent states}\label{subcoh}
Before we discuss how to use the framework in the previous section to study the complexity in TFD states, let's first try to study a useful model in quantum mechanics. As lots of new definitions and clarifications were made in previous sections, it is better to use some simple examples to familiarize the readers with them. In this subsection, let's assume that the momentum has only one possible value so that we can neglect the momentum index. Then the Hilbert space is spanned by $\{|n\rangle\}$. We choose that states set $\mathcal{S}_E$ is the collection of all the coherent states,
\begin{equation}\label{cherent1}
  \mathcal{S}_E:=\{|\coh(\alpha)\rangle|~\forall|\coh(\alpha)\rangle\in\mathcal{H}, s.t., \hat{a}|\coh(\alpha)\rangle=\alpha|\coh(\alpha)\rangle\}\,.
\end{equation}
We see that $\mathcal{S}_E$ is the collection of all the eigenstates of lowering/annihilation operator\footnote{All the results in this subsection can be used into the both of free and interacted systems.}. This state can be generated  from vacuum state $|0\rangle$ by displacement operator $\hat{D}_\alpha:=\exp(\alpha\hat{a}^\dagger-\alpha^*\hat{a})$,
\begin{equation}\label{cherent2}
  |\coh(\alpha)\rangle=\hat{D}_\alpha|0\rangle\,.
\end{equation}
It is obvious that displacement operators are the elements of $\mathcal{U}_E$. In order to use our method to compute the complexity between any two states in $\mathcal{S}_E$, we have to check that if there is at least one operator in $\mathcal{U}_E$ to convert each other of any two states in $\mathcal{S}_E$. This can be done as follows. Firstly, one can prove that the displacement operator $\hat{D}_\alpha$ and $\hat{D}_\beta$ satisfy,
\begin{equation}\label{DaDb1}
  \hat{D}_\alpha \hat{D}_\beta=e^{\alpha\beta^*-\beta\alpha^*}\hat{D}_{\alpha+\beta}\,.
\end{equation}
This equation implies that $\hat{D}_\alpha|\coh(\beta)\rangle=e^{\alpha\beta^*-\beta\alpha^*}|\coh(\alpha+\beta)\rangle\sim|\coh(\alpha+\beta)\rangle$, so any two elements in set~\eqref{cherent1} can be converted to each other by at leas one operator in $\mathcal{U}_E$.

Now let's try to compute the complexity from $|0\rangle$ to $|\coh(\alpha)\rangle$ according to Eq.~\eqref{defcompstates}. To do that we have to find all the operators $\hat{U}$ such that $|\coh(\alpha)\rangle\sim \hat{U}|0\rangle$. Displacement operator $\hat{D}_\alpha$ of course is one of such operators but is not the one of minimal complexity. In fact, all operators $\hat{U}_f$ with the form of $\exp[\alpha\hat{a}^\dagger+f\hat{a}]$ for arbitrary constant $f$ can satisfy that $|\coh(\alpha)\rangle\sim\hat{U}_f|0\rangle$.

Let's first show how to compute the complexity of $\hat{U}_0=\exp[\alpha\hat{a}^\dagger]$.
The general curve in $\mathcal{U}_E$ is generated by following generator,
\begin{equation}\label{generatorcoh1}
  \hat{T}(t)=T_0(t)\hat{\mathbb{I}}+Y_1(t)\hat{a}^++Y_2(t)\hat{a}+\sum_{i_1,i_2}Y_{i_1i_2}\hat{b}_{i_1}\hat{b}_{i_2}+\cdots+\sum_{{i_1},\cdots,{i_n}}Y_{i_1\cdots i_n}\hat{b}_{i_1}\cdots\hat{b}_{i_n}+\cdots
\end{equation}
Here $\hat{b}_{i_1},\cdots,\hat{b}_{i_n}\in \{\hat{a}, \hat{a}^\dagger\}$. As the curve generated by $\hat{T}(t)$ should satisfy condition $\hat{U}_0=\exp[\alpha\hat{a}^\dagger]=\hat{c}(1)=\lambda\overleftarrow{\mathcal{P}}\exp\int_0^1\hat{T}(t)\td t$, so we obtain the restricted extremum problem,
\begin{equation}\label{Complxity-coh}
  \mathcal{C}(\hat{U}_0)=\min\left\{\int_0^1\td t\left[\parallel Y_1(t)\parallel+\parallel Y_2(t)\parallel +2\sum_{i_1,i_2}\parallel Y_{i_1i_2}\parallel+\cdots+\cdots\right]\right\}
\end{equation}
with the constraint,
\begin{equation}\label{constaint1}
\begin{split}
  \exp(\alpha\hat{a}^\dagger)=&\lambda\overleftarrow{P}\exp\left\{\int_0^1\td t\left[T_0(t)\hat{\mathbb{I}}+Y_1(t)\hat{a}^++Y_2(t)\hat{a}+\sum_{i_1,i_2}Y_{i_1i_2}\hat{b}_{i_1}\hat{b}_{i_2}+\right.\right.\\
  &\left.\left.\cdots+\sum_{{i_1},\cdots,{i_n}}Y_{i_1\cdots i_n}\hat{b}_{i_1}\cdots\hat{b}_{i_n}+\cdots\right]\right\}\,
  \end{split}
\end{equation}
for a nonzero complex number $\lambda$.

It seems to be a high challenge to solve optimization problem Eqs.~\eqref{Complxity-coh} and \eqref{constaint1} strictly. As the first attempt to investigate the complexity in this manner, in order to avoid to sink into verbose math, let's reduce the elements in generators set. Here it is assumed that
\begin{equation}\label{reduceE}
  E=E^{(0)}=\{\hat{a},\hat{a}^\dagger,\hat{\mathbb{I}}\}\,.
\end{equation}
Under this reduced generators set, the optimization problem Eqs.~\eqref{Complxity-coh} and \eqref{constaint1} then becomes,
\begin{equation}\label{Complxity-coh2}
  \mathcal{C}(U_0)=\min\left\{\int_0^1\td t\left[\parallel Y_1(t)\parallel+\parallel Y_2(t)\parallel \right]\right\}
\end{equation}
with the constraint,
\begin{equation}\label{constaint1b}
\begin{split}
  \exp(\alpha\hat{a}^\dagger)=&\lambda\overleftarrow{P}\exp\left\{\int_0^1\td t\left[T_0(t)\hat{\mathbb{I}}+Y_1(t)\hat{a}^++Y_2(t)\hat{a}\right]\right\}\,
  \end{split}
\end{equation}
for a nonzero complex number $\lambda$.
Using the results in the appendix~\ref{coherent1}, we can find that the complexity of $\hat{U}_0$ is,
\begin{equation}\label{complexityU00}
  \mathcal{C}(\hat{U}_0)=\mathcal{C}(\exp[\alpha\hat{a}^\dagger])=\parallel\alpha\parallel\,.
\end{equation}
Here $\parallel\alpha\parallel :=\rho(|\cos\theta|+|\sin\theta|\cdot|\theta|)$ for $\alpha=\rho e^{i\theta}$ and $\theta\in[-\pi,\pi)$. The complexity between the coherent state and vacuum state is (see Eq.~\eqref{Ccohb1b2} in appendix~\ref{coherent1}),
\begin{equation}\label{compl-coh2s}
  \mathcal{C}(|\coh(\alpha)\rangle,|0\rangle)=\parallel\alpha\parallel \,.
\end{equation}
If someone directly uses the complexity of displacement operator $\hat{D}_\alpha$ to stand for the complexity $|0\rangle\rightarrow|\coh(\alpha)\rangle$ then he will find that its a value is $2\parallel\alpha\parallel$, which is larger than the result in Eq.~\eqref{compl-coh2s}. Of course, if one insist that the operators we can use to convert states are unitary, then displacement operator $\hat{D}_\alpha$ is  one of which give the minimal complexity. Just as mentioned in the introduction part, it may lead to larger complexity by reducing the operator sets.

Using the relationship $\exp(\beta\hat{a}^\dagger)\exp(\alpha\hat{a}^\dagger)=\exp[(\beta+\alpha)\hat{a}^\dagger]$, we can see that $$\exp(-\alpha\hat{a}^\dagger) |\coh(\alpha)\rangle=|0\rangle\,,$$ so we have,

\begin{equation}\label{compl-coh3}
  \mathcal{C}(|0\rangle,|\coh(\alpha)\rangle)=\mathcal{C}(\exp[-\alpha\hat{a}^\dagger])=\parallel\alpha\parallel \,.
\end{equation}
and,
\begin{equation}\label{compl-coh40}
  \mathcal{C}(|\coh(\beta)\rangle,|\coh(\alpha)\rangle)=\parallel\beta-\alpha\parallel \,.
\end{equation}
We see that in this case the complexity between coherent states has reversibility. By these results we can check the properties (1b) and (2b) in the Sec.~\ref{Intro},
\begin{equation}\label{two-prop1}
\begin{split}
  &\mathcal{C}(|\coh(\beta)\rangle,|\coh(\alpha)\rangle)=0\Leftrightarrow\alpha=\beta\Leftrightarrow|\coh(\beta)\rangle\sim|\coh(\alpha)\rangle\\
  &\mathcal{C}(|\coh(\beta)\rangle,|\coh(\alpha)\rangle)+\mathcal{C}(|\coh(\alpha)\rangle,|0\rangle)\geq\mathcal{C}(|\coh(\beta)\rangle,|0\rangle)\,.
  \end{split}
\end{equation}
For any reference coherent state $|\coh(\gamma)\rangle$,  we have following inequality for the complexity of formation,
\begin{equation}\label{cohcomfor}
  \Delta\mathcal{C}_\gamma(|\coh(\beta)\rangle,|\coh(\alpha)\rangle)=\parallel\beta-\gamma\parallel -\parallel\alpha-\gamma\parallel \leq\parallel\alpha-\beta\parallel \,.
\end{equation}
We see that the complexity of formation depends in the choice of reference state $|\coh(\gamma)\rangle$.

When we recover the generators set $E$ into the form in Eq.~\eqref{extendbasis}, it seems that the complexity between coherent state and vacuum state is still given by Eq.~\eqref{complexityU00}. The proof is not obtained yet but the physical intuition for such predication is simple: there is no any ladder operators to be wasted (the meaning of ``wasted'' here is that a particle created/annihilated at earlier time will be annihilated/created at the later time), so it contains the minimal operators to convert the reference state into the target state.

\subsection{Complexity of entangled thermal states}\label{ETS1}
In this subsection, we still restrict the consideration in the case that there are two kinds of created and annihilated  operators. By this subsection, we want to show and clarity the similarity and differences between the complexity and other conceptions such as thermal/entanglement entropy and complexity of formation.

Let's consider a Hilbert space $\mathcal{H}=\mathcal{H}_1\times\mathcal{H}_2$ so we have two groups of creation and annihilation  operators and $E^0:=\{\hat{a}_1,\hat{a}_2,\hat{a}_1^\dagger,\hat{a}_2^\dagger\}$. Let's consider the entangled thermal state,\footnote{Similar to the previous subsection, here it is still not assumed that the system is free system.}
\begin{equation}\label{ETS}
  |S(\beta)\rangle:=\sqrt{1-e^{-\beta\omega}}\sum_{n=0}^{\infty}e^{-\beta n\omega/2}|n\rangle_1|n\rangle_2\,.
\end{equation}
Here $\omega>0$ is the energy of every one particle. The normalization constant has been added so that $\langle S(\beta)|S(\beta)\rangle=1$. The density matrix is,
\begin{equation}\label{densityM1}
  \rho:=|S(\beta)\rangle\langle S(\beta)|=(1-e^{-\beta\omega})\sum_{n,m=0}^{\infty}e^{-\beta (n+m)\omega/2}|n\rangle_1|n\rangle_2\langle m|_1\langle m|_2\,.
\end{equation}
As $\rho$ is the density matrix for pure state, the thermal entropy of this system is zero. In order to find the entanglement entropy between the subspace $\mathcal{H}_1$ and $\mathcal{H}_2$, let's first take the trace of $\mathcal{H}_2$ in the density matrix,
\begin{equation}\label{densityM2}
  \rho_1=\text{Tr}_2(\rho)=\sum_{m=0}^\infty\langle m|{_2}\rho|m\rangle_2=\frac{1}{Z}\sum_{n=0}^{\infty}e^{-\beta n\omega}|n\rangle_1\langle n|_1\,.
\end{equation}
Then we see that the subsystem is a mix state system with temperature $T=1/\beta$. We can read the partition function $Z(\beta)=1/(1-e^{-\beta\omega})$ and the entanglement entropy $S_{12}$ is,
\begin{equation}\label{partittionZ1}
S_{12}=-\text{Tr}(\rho_1\ln\rho_1)=\ln Z-\beta\frac{\partial}{\partial\beta}\ln Z=-\ln(1-e^{-\beta\omega})+\frac{\beta\omega}{e^{\beta\omega}-1}\,.
\end{equation}

Now let's try to compute the complexity between $|S(\beta)\rangle$ and its corresponding ground state. To do so, let's written the state $|S(\beta)\rangle$ as follows,
\begin{equation}\label{ETS2}
   |S(\beta)\rangle\sim\sum_{n=0}^{\infty}\frac{e^{-\beta n\omega/2}}{n!}(\hat{a}_1^\dagger\hat{a}_2^\dagger)^n|0\rangle_1|0\rangle_2=\hat{U}_\beta|0\rangle_1|0\rangle_2\,.
\end{equation}
Here $\hat{U}_\beta:=\exp(e^{-\beta\omega/2}\hat{a}_1^\dagger\hat{a}_2^\dagger)$. In fact $|S(\beta)\rangle$ can be regarded as the TFD state in quantum mechanics ((1+0)-dimensional quantum field). We here consider the complexity of conversion $|0\rangle_1|0\rangle_2\rightarrow|S(\beta)\rangle$.
It seems a high challenge to find the complexity under the generators set \eqref{extendbasis}. Let's assume that the generators set only contains two elements $$E=\{\hat{a}_1^\dagger\hat{a}_2^\dagger,\hat{\mathbb{I}}\}.$$
Then any operator $\hat{U}\in\mathcal{U}_E$ has the relationship $\hat{U}\sim\exp(\lambda\hat{a}_1^\dagger\hat{a}_2^\dagger)$ with $\lambda\in\mathbb{C}$. To determine the complexity of $\hat{U}$, we have to solve the restricted extremum problem,
\begin{equation}\label{Complxity-ETS}
  \mathcal{C}(\hat{U})=2\min\left\{\int_0^1\td t\parallel Y_1(t)\parallel\right\}
\end{equation}
with the constraint,
\begin{equation}\label{constaint2}
\begin{split}
  \hat{U}\sim\exp(\lambda\hat{a}_1^\dagger\hat{a}_2^\dagger)\sim\overleftarrow{P}\exp\left\{\int_0^1\td tY_{1}(t)\hat{a}_1^\dagger\hat{a}_2^\dagger\right\}=\exp\left\{\hat{a}_1^\dagger\hat{a}_2^\dagger\int_0^1\td tY_{1}(t)\right\}\,.
  \end{split}
\end{equation}
Solving this optimization problem, we can find that
$$\mathcal{C}[\exp(\lambda\hat{a}_1^\dagger\hat{a}_2^\dagger)]=2\parallel \lambda\parallel .$$
Hence, the complexity of $\hat{U}_\beta$ is,
\begin{equation}\label{complexityU0}
  \mathcal{C}(\hat{U}_\beta)=2e^{-\beta\omega/2}\,
\end{equation}
All the operators which can transform $|0\rangle_1|0\rangle_2$ into $|S(\beta)\rangle$ are equivalent to $\hat{U}_\beta$, so  we obtain that,
\begin{equation}\label{compl-coh2}
  \mathcal{C}(|S(\beta)\rangle,|0\rangle_1|0\rangle_2)=2e^{-\beta\omega/2}\,.
\end{equation}
We see that in general $\mathcal{C}(|S(\beta)\rangle,|0\rangle_1|0\rangle_2)\neq S_{12}$. This shows that complexity and entanglement entropy are different quantities. In fact, the complexity is the defined between two states, so we can change the reference state and compute the complexity between $|S(\beta)\rangle$ and this reference state. Then the value of complexity in general has no direct relationship to entanglement entropy. This shows that complexity in fact is a new independent quantity to describe the relationship between two states.


One can easy see that $\exp(e^{-\beta_1\omega/2}\hat{a}_1^\dagger\hat{a}_2^\dagger)\exp(e^{-\beta_2\omega/2}\hat{a}_1^\dagger\hat{a}_2^\dagger)=\exp[(e^{-\beta_1\omega/2}+e^{-\beta_2\omega/2})\hat{a}_1^\dagger\hat{a}_2^\dagger]$, so we have,
\begin{equation}\label{twosbetas}
  \mathcal{C}(|S(\beta_2)\rangle,|S(\beta_1)\rangle)=2|e^{-\beta_2\omega/2}-e^{-\beta_1\omega/2}|\,.
\end{equation}
The reversible condition is also satisfied in this case.  Eq.~\eqref{twosbetas} leads that the complexity of formation of $|S(\beta_1)\rangle$ and $|S(\beta_2)\rangle$ corresponding to $|S(\beta_3)\rangle$ is,
\begin{equation}\label{compl-coh3}
  \Delta\mathcal{C}_{\beta_3}(|S(\beta_2)\rangle,|S(\beta_1)\rangle)=2|e^{-\frac{\beta_2\omega}2}-e^{-\frac{\beta_3\omega}2}|-2|e^{-\frac{\beta_3\omega}2}-e^{-\frac{\beta_1\omega}2}| \leq\mathcal{C}(|S(\beta_2)\rangle,|S(\beta_1)\rangle)\,.
\end{equation}
Specially, when $\beta_3\leq\min\{\beta_2, \beta_1\}$ or $\beta_3\geq\max\{\beta_2, \beta_1\}$, the left-hand of Eq.~\eqref{compl-coh3} is independent of the value of $\beta_3$ and the absolute value of complexity of formation is just the complexity between these two states,
\begin{equation}\label{compl-coh4}
  \mathcal{C}(|S(\beta_2)\rangle,|S(\beta_1)\rangle)=|\Delta\mathcal{C}_{\beta_3}(|S(\beta_2)\rangle,|S(\beta_1)\rangle)|\,.
\end{equation}
This can be understood physically by following argument.
\begin{figure}
  \centering
  \includegraphics[width=.4\textwidth]{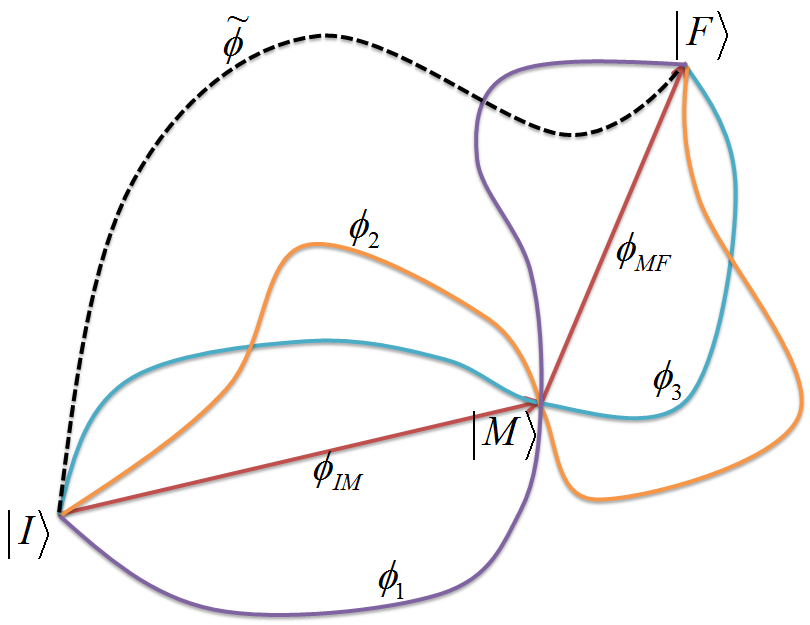}
  \caption{Curves $\phi_{IM},\phi_{MF}, \phi_1, \phi_2, \phi_3,\cdots$ stand for the possible quantum circuits  which can covert the initial state $|I\rangle$ into the finial state $|F\rangle$. The medial state $|M\rangle$ is the necessary state that all the physically realizable quantum circuit will bring the initial state $|I\rangle$ into the medial state $|M\rangle$ before it reaches the finial state $|F\rangle$. The black dashed curve $\tilde{\phi}$ stands for a quantum circuit which can connect states $|I\rangle$ and $|F\rangle$ without passing through the medial state $|M\rangle$. But this curve is forbidden by some physical rules. }\label{ineqs2}
\end{figure}

Let's consider to design some quantum circuits to covert the initial state $|I\rangle$ into the finial state $|F\rangle$.  One can see Fig.~\ref{ineqs2}. Different curves such as $\phi_{IM},\phi_{MF}, \phi_1, \phi_2, \phi_3,\cdots$ stand for different designs. In general, some designs will bring the initial state passing through the medial state $|M\rangle$ but some designs such as $\tilde{\phi}$ will not. However, in some special cases, because of physical restrictions, all the physical realizable quantum circuits will bring the initial state passing through the medial state $|M\rangle$. In these cases, if $\phi_{IM}$ is one quantum circuit of minimal gates which can convert $|I\rangle$ to $|M\rangle$ and $\phi_{MF}$ is one quantum circuit of minimal gates which can convert $|M\rangle$ to $|F\rangle$, then the combination $\phi_{MF}\circ\phi_{IM}$ is also one quantum circuit of minimal gates to realize the conversion $|I\rangle$ to $|F\rangle$. This means that,
\begin{equation}\label{complIMF1}
  \mathcal{C}(|F\rangle,|I\rangle)=\mathcal{C}(|F\rangle,|M\rangle)+\mathcal{C}(|M\rangle,|I\rangle)\,,
\end{equation}
or,
\begin{equation}\label{complIMF2}
  \Delta\mathcal{C}_I(|F\rangle,|M\rangle):=\mathcal{C}(|F\rangle,|I\rangle)-\mathcal{C}(|M\rangle,|I\rangle)=\mathcal{C}(|F\rangle,|M\rangle)\,,
\end{equation}
In addition, if the complexity is reversible, then Eq.~\eqref{complIMF2} can be strengthened as,
\begin{equation}\label{complIMF3}
  |\Delta\mathcal{C}_I(|F\rangle,|M\rangle)|=\mathcal{C}(|F\rangle,|M\rangle)\,,
\end{equation}
We see that the absolute value of complexity of formation between states $|F\rangle$ and $|M\rangle$ (corresponding to $|I\rangle$) then is just the complexity between them.

Now let's return to the case in  entangled thermal states. Let's assume $\beta_1\leq\beta_2\leq\beta_3$ and consider the conversion $|S(\beta_3)\rangle\rightarrow|S(\beta_1)\rangle$. As the parameter $\beta$ in the states $|S(\beta)\rangle$ describe the temperature and can only be changed continuously in a real system, then all the physically realizable quantum circuits must bring the state $|S(\beta_3)\rangle$ into the state $|S(\beta_2)\rangle$ before then reach the finial state $|S(\beta_1)\rangle$. As the complexity satisfies reversibility in entangled thermal states, we see that,
\begin{equation}\label{complIMF4}
  |\Delta\mathcal{C}_{\beta_3}(|S(\beta_1)\rangle,|S(\beta_2)\rangle)|=\mathcal{C}(|S(\beta_1)\rangle,|S(\beta_2)\rangle)\,,
\end{equation}
For other cases of $\beta_3\leq\min\{\beta_2, \beta_1\}$ or $\beta_3\geq\max\{\beta_2, \beta_1\}$, one can find the same result.

\section{Complexity between TFD states}\label{Com-TFD}
\subsection{Construct TFD states  by Bogoliubov transformations}\label{TFD2}
In this section, we will construct two different vacuum states which can be associated by a Bogoliubov transformation. One can see that the TFD state can be naturally identified with a vacuum state by this manner. For simplicity, we only consider the free scalar field theory in this section.

Let's first define a two-copy vacuum state $|A\rangle:=|A\rangle_L |A\rangle_R$. The state $|A\rangle_L$ is left-side vacuum state in the Hilbert space $\mathcal{H}_L$, which is annihilated by the operator $\hat{a}_{\vec{k}}^{L}$,
\begin{equation}\label{vacL1}
  \hat{a}_{\vec{k}}^{L}|A\rangle_L=0\,.
\end{equation}
Here $\vec{k}=(k^1,k^2,\cdots,k^{d-1})$ is the momentum of the annihilated particle. The state $|A\rangle_R$ is right-side vacuum state in the Hilbert space $\mathcal{H}_R$ which is the copy of $\mathcal{H}_L$. Its annihilation operator is $\hat{a}_{\vec{k}}^{R}$. Let's first assume that the momentum is discrete. Then the annihilation and creation operators corresponding to the left-side and right-side vacuum states satisfy following bosonic commutation relations,
\begin{equation}\label{comuubos}
  [ \hat{a}_{\vec{k}}^{L}, \hat{a}_{\vec{k}'}^{L\dagger}]= [ \hat{a}_{\vec{k}}^{R}, \hat{a}_{\vec{k}'}^{R\dagger}]=\delta_{\vec{k}\vec{k}'}\hat{\mathbb{I}}
\end{equation}
and others are zeros. All the excited states in the Hilbert space $\mathcal{H}:=\mathcal{H}_L\times\mathcal{H}_R$ then can be generated by using creation operators $\hat{a}_{\vec{k}}^{L\dagger}$ and $\hat{a}_{\vec{k}}^{R\dagger}$ from vacuum state $|A\rangle$.

The vacuum state is not the unique. In fact, the studies on the quantum field in curved spacetime have made us to realize that vacuum also depends on the observers. This understanding leads to the unified understanding on the Unruh effect~\cite{Unruh1976,Crispino:2007eb}, Hawking radiation~\cite{HAWKING1974,hawking1975} and other particle automatical creations in the curved spacetime~\cite{birrell1982quantum,Jacobson:2003vx}. To define an other vacuum state, let's consider a new decomposition on the Hilbert space $\mathcal{H}$ such that $\mathcal{H}=\mathcal{H}_U\times\mathcal{H}_D$ and the corresponding annihilation operators $(\hat{b}_{\vec{k}}^{U},\hat{b}_{\vec{k}}^{D})$ which have following relationships to $(\hat{a}_{\vec{k}}^{L},\hat{a}_{\vec{k}}^{R})$,
\begin{equation}\label{bogoliu1}
  \hat{b}_{\vec{k}}^{U}:=c_{\vec{k}}(\hat{a}_{\vec{k}}^{R}-e^{-\pi\omega_{\vec{k}}/a}\hat{a}_{\vec{k}}^{L\dagger}), ~~\hat{b}_{\vec{k}}^{D}:=d_{\vec{k}}(\hat{a}_{\vec{k}}^{L}-e^{-\pi\omega_{\vec{k}}/a}\hat{a}_{\vec{k}}^{R\dagger})\,.
\end{equation}
Here $\omega_{\vec{k}}$ is the energy of a particle at momentum $\vec{k}$ and $a$ is a non-negative real number. We will see later on that $a$ is just proportional to the temperature of a TFD state. For the free scalar field with conformal symmetry, the mass is zero and we have $\omega_{\vec{k}}=\sqrt{\vec{k}^2}$. The coefficients $c_{\vec{k}}$ and $d_{\vec{k}}$ are determined so that bosonic commutation relationships are satisfied,
\begin{equation}\label{comuubos}
  [ \hat{b}_{\vec{k}}^{U}, \hat{b}_{\vec{k}'}^{U\dagger}]= [ \hat{b}_{\vec{k}}^{D}, \hat{b}_{\vec{k}'}^{D\dagger}]=\delta_{\vec{k}\vec{k}'}\hat{\mathbb{I}}\,.
\end{equation}
As they are not important in this paper, we will not give out their expressions. The annihilation operators pairs $(\hat{b}_{\vec{k}}^{U},\hat{b}_{\vec{k}}^{D})$ determine a new vacuum $|B\rangle:=|B\rangle_{U} |B\rangle_{D}$, which satisfies,
\begin{equation}\label{vacU1}
  \hat{b}_{\vec{k}}^{U}|B\rangle_U=\hat{b}_{\vec{k}}^{D}|B\rangle_D=0\,,
\end{equation}
or we can write that $\hat{b}_{\vec{k}}^{U}|B\rangle=\hat{b}_{\vec{k}}^{D}|B\rangle=0$.

We see that in the Eq.~\eqref{bogoliu1} the new annihilation operators are mixed with original creation and annihilation operators, so from the viewpoint of vacuum $|A\rangle$, the state $|B\rangle$ is an excited state and has nonzero particle numbers. Using the Eqs.~\eqref{bogoliu1} and \eqref{vacU1}, one can easy find following relationship,
\begin{equation}\label{vacrel1}
  (\hat{a}_{\vec{k}}^{R\dagger}\hat{a}_{\vec{k}}^{R}-\hat{a}_{\vec{k}}^{L\dagger}\hat{a}_{\vec{k}}^{L})|B\rangle=0
\end{equation}
This equation implies that the number of left-side particles and the number of right-side particles are the same in the new vacuum state $|B\rangle$. As the Fock's space of $\mathcal{H}_L$ and $\mathcal{H}_R$ can span the Hilbert space $\mathcal{H}$, any state in $\mathcal{H}$ can be written as the superposition of the particle states in them. Then we see that,
\begin{equation}\label{TFDdep1}
  |B\rangle\sim\left[\prod_{\vec{k}_i}\sum_{n=0}^{\infty}\frac{K_{n}}{n!}(\hat{a}_{\vec{k}_i}^{R\dagger} \hat{a}_{\vec{k}_i}^{L\dagger})^{n_i}\right]|A\rangle\,.
\end{equation}
The recursion formula for the coefficients $K_{n_i}$ can be readily found from the Eq.~\eqref{vacrel1}. The result is,
\begin{equation}\label{recurK}
  K_{n+1}-e^{-\pi\omega_{\vec{k}_i}/a}K_{n}=0\,.
\end{equation}
Then we can see that,
\begin{equation}\label{TFDdep2}
  |B\rangle\sim\left[\prod_{\vec{k}_i}\sum_{n=0}^{\infty}\frac{e^{-\pi n_i\omega_{\vec{k}_i}/a}}{n!}(\hat{a}_{\vec{k}_i}^{R\dagger} \hat{a}_{\vec{k}_i}^{L\dagger})^{n}\right]|A\rangle =\prod_{\vec{k}_i}\sum_{n=0}^{\infty}e^{-\pi n\omega_{\vec{k}_i}/a}|n,\vec{k}_i\rangle_L |n,\vec{k}_i\rangle_R\,.
\end{equation}
%
Here the notation $|n,\vec{k}_i\rangle$ means that there are $n$ particles with momentum $\vec{k}_i$.
The similar relation between state $|B\rangle$ and $|A\rangle$ appears in some important physical situations. For example, for an accelerated observer in Minkowskin spacetime, the vacuum $|A\rangle$ is the Rindler vacuum and the vacuum $|B\rangle$ is the  Minkowskin vacuum, Eq.~\eqref{TFDdep2} then shows that accelerated observer can find the particles appearing in Minkowskin vacuum, which leads to the Unruh effect. For the a static  observer at the infinite of a Schwarzschild black hole, the vacuum $|A\rangle$ is the out-vacuum and the vacuum $|B\rangle$ is in-vacuum, Eq.~\eqref{TFDdep2} then leads to the particles emission from the black hole, which is the origin of Hawking radiation.

In order to see that the vacuum state corresponding to annihilation operators pair $(\hat{b}_{\vec{k}}^{U},\hat{b}_{\vec{k}}^{D})$ is just a TFD state, let's introduce the energy eigenstate $|E_i\rangle_L$ and $|E_i\rangle_R$ as the basis of Hilbert space $\mathcal{H}_L$ and $\mathcal{H}_R$, then the state $|B\rangle$ can be expressed as,
\begin{equation}\label{BexpbyE1}
  |B\rangle\sim \sum_{i,j=0}^\infty f_{ij}|E_i\rangle_L |E_j\rangle_R\,.
\end{equation}
For every momentum $\vec{k}_i$, the left and right sides always contain the same particles, which means that they also have the same energy. For every state with momentum $\vec{k}_i$ and particle number $n$, it contains the energy $E=n\omega_{\vec{k}_i}$ and the coefficient with is probational to  $e^{-\pi n\omega_{\vec{k}_i}/a}$. This means that $f_{ij}=\delta_{ij}\exp(-\pi E_i/a)$ and so,
\begin{equation}\label{BexpbyE2}
  |B\rangle\sim \sum_{i=0}^\infty e^{-\pi E_i/a}|E_i\rangle_L |E_i\rangle_R\,.
\end{equation}
Here $|E_i\rangle_{L/R}:=g(E_i)\prod_{\vec{k}_j}\sum_{n_i}|n_i,\vec{k}_j\rangle_{L/R}$ with the restriction $\omega_{\vec{k}_i}\sum n_i=E_i$. Here the coefficient $g(E_i)$ is the normalization factor. Comparing it with the Eq.~\eqref{TFD1}, one can see that $|B\rangle$ is a TFD state with temperature $T=a/2\pi$. The vacuum state $|A\rangle$ is just the TFD state at the zero temperature limit,.i.e., the state at the limit $a\rightarrow0$.

As all the different $\hat{a}_{\vec{k}_i}^{R\dagger} \hat{a}_{\vec{k}_i}^{L\dagger}$ are commutative, the Eq.~\eqref{TFDdep2} can be written into the continuous form by this way,
\begin{equation}\label{TFDdep3}
\begin{split}
  |\text{TFD}\rangle:=&|B\rangle\sim\prod_{\vec{k}_i}\sum_{n_i=0}^{\infty}\frac{e^{-\pi n_i\omega_{\vec{k}_i}/a}}{n_i!}(\hat{a}_{\vec{k}_i}^{R\dagger} \hat{a}_{\vec{k}_i}^{L\dagger})^{n_i}|A\rangle =\prod_{\vec{k}_i}\exp[e^{-\pi\omega_{\vec{k}_i}/a}\hat{a}_{\vec{k}_i}^{R\dagger} \hat{a}_{\vec{k}_i}^{L\dagger}]|A\rangle\\
  &=\exp\left[\sum_{\vec{k}_i}e^{-\pi\omega_{\vec{k}_i}/a}\hat{a}^{R\dagger}_{\vec{k}_i} \hat{a}^{L\dagger}_{\vec{k}_i}\right]|A\rangle\\
  &=\exp\left[\int\td k^{d-1}e^{-\pi\omega_{\vec{k}}/a}\hat{a}^{R\dagger}(\vec{k}) \hat{a}^{L\dagger}(\vec{k})\right]|A\rangle\\
  &=\hat{U}_a^\dagger|A\rangle
  \end{split}
\end{equation}
with the operator $\hat{U}_a^\dagger$ defined as,
\begin{equation}\label{defUa}
  \hat{U}_a^\dagger:=\exp\left[\int\td k^{d-1}e^{-\pi\omega_{\vec{k}}/a}\hat{a}^{R\dagger}(\vec{k}) \hat{a}^{L\dagger}(\vec{k})\right]\,.
\end{equation}
The discrete creation operator $\hat{a}_{\vec{k}_i}^\dagger$ has been converted into its continuous form by Eqs.~\eqref{adtoac} and \eqref{sumint}.

The continuous form in Eq.~\eqref{TFDdep3} shows that a TFD state and its vacuum can be associated by the operator $\hat{U}_a^\dagger$. This is the starting point in following computations about complexity. A remarkable property  is that we do not need any UV cut-off at the momentum when we construct the TFD state from the vacuum state by Bogoliubov transformation. It needs to note that the operator $\hat{U}_a^\dagger$ is not unitary as we can see that $\hat{U}_a^{-1}\neq\hat{U}_a^\dagger$. However, $\hat{U}_a^\dagger$ has a unitary partner $\hat{G}_a$ which can also realize the conversation from $|A\rangle$ to $|\text{TFD}\rangle$,
\begin{equation}\label{defineG1}
  \hat{G}_a:=\exp\left\{\int\td k^{d-1}f(a,\vec{k})[\hat{a}^{R\dagger}(\vec{k}) \hat{a}^{L\dagger}(\vec{k})-\hat{a}^{R}(\vec{k}) \hat{a}^{L}(\vec{k})]\right\}
\end{equation}
for a real-valued function $f(a,\vec{k})$. One can see that $\hat{G}_a$ is unitary as $\hat{G}_a\hat{G}_a^\dagger=\hat{I}$. Let's try to find a function $f(a,\vec{k})$ so that $\hat{U}_a^\dagger|A\rangle=\hat{G}_a|A\rangle$. Firstly, we return to the discrete form and define,
\begin{equation}\label{generatL}
\begin{split}
  &\hat{L}_+^{(\vec{k})}:={\hat{a}^{R\dagger}_{\vec{k}} \hat{a}^{L\dagger}_{\vec{k}}},~~\hat{L}_-^{(\vec{k})}={\hat{a}^{R}_{\vec{k}} \hat{a}^{L}_{\vec{k}}}, \\
  &\hat{L}_0^{(\vec{k})}:=\frac12(\hat{a}^{R}_{\vec{k}}\hat{a}^{R\dagger}_{\vec{k}}+\hat{a}^{L}_{\vec{k}} \hat{a}^{L\dagger}_{\vec{k}}-\hat{\mathbb{I}})=\frac12(\hat{a}^{R\dagger}_{\vec{k}}\hat{a}^{R}_{\vec{k}}+ \hat{a}^{L\dagger}_{\vec{k}}\hat{a}^{L}_{\vec{k}}+\hat{\mathbb{I}}),\\
  &\hat{G}_a^{\vec{k}}:=\exp\left[f(a,\vec{k})(\hat{L}_+^{(\vec{k})}-\hat{L}_-^{(\vec{k})})\right]\,.
  \end{split}
\end{equation}
Then Eqs.~\eqref{defineG1} and \eqref{TFDdep3} can be presented as,
\begin{equation}\label{definGLs}
  \hat{G}_a=\prod_{\vec{k}}\exp\left[f(a,\vec{k})(\hat{L}_+^{(\vec{k})}-\hat{L}_-^{(\vec{k})})\right]=\prod_{\vec{k}}\hat{G}_a^{\vec{k}}\,,
\end{equation}
\begin{equation}\label{TFDdep3b}
  |\text{TFD}\rangle\sim\prod_{\vec{k}}\exp\left[e^{-\pi\omega_{\vec{k}}/a}\hat{L}_+^{(\vec{k})}\right]|A\rangle\,.
\end{equation}
The generators $\{\hat{L}_+^{(\vec{k})},\hat{L}_-^{(\vec{k})},\hat{L}_0^{(\vec{k})}\}$ form a su(1,1) Lie-algebra with the commutation relationships,
\begin{equation}\label{commtrel}
  [\hat{L}_+^{(\vec{k})},\hat{L}_-^{(\vec{k})}]=-2\hat{L}_0^{(\vec{k})},~~~[L_0^{(\vec{k})},\hat{L}_\pm^{(\vec{k})}]=\pm\hat{L}_\pm^{(\vec{k})}\,.
\end{equation}
$\hat{G}_a$ defined in Eq.~\eqref{generatL} can be decomposed as~(see the appendix 11.3.3 of Ref.~\cite{klimov2009a}),
\begin{equation}\label{decompGak}
  \hat{G}_a^{\vec{k}}=\exp\left[\gamma_+(\vec{k})\hat{L}_+^{(\vec{k})}\right]\exp\left[\ln\gamma_0(\vec{k})\hat{L}_0^{(\vec{k})}\right]\exp\left[\gamma_-(\vec{k})\hat{L}_-^{(\vec{k})}\right]
\end{equation}
with
\begin{equation}\label{gammapm0}
  \gamma_\pm(\vec{k})=\pm\tanh f(a,\vec{k}),~~\gamma_0(\vec{k})=\cosh^{-2}f(a,\vec{k})\,.
\end{equation}
Noting the fact that $\hat{L}_-^{(\vec{k})}|A\rangle=0$ and $\hat{L}_0^{(\vec{k})}|A\rangle=\frac12|A\rangle$, one can find that,
\begin{equation}\label{decompGak2}
  \hat{G}_a^{\vec{k}}|A\rangle=\exp\left[\gamma_+(\vec{k})\hat{L}_+^{(\vec{k})}\right]\exp\left[\frac12\ln\gamma_0(\vec{k})\right]|A\rangle\sim\exp\left[\gamma_+(\vec{k})\hat{L}_+^{(\vec{k})}\right]|A\rangle\,.
\end{equation}
So we see that $\hat{G}_a|A\rangle\sim\prod_{\vec{k}}\exp\left[\gamma_+(\vec{k})\hat{L}_+^{(\vec{k})}\right]|A\rangle$. Comparing it with Eq.~\eqref{TFDdep3b}, we see that $\hat{G}_a|A\rangle=\hat{U}_a^\dagger|A\rangle$ if we take,
\begin{equation}\label{fak2}
  f(a,\vec{k})=\text{arctanh}e^{-\pi\omega_{\vec{k}}/a}\,.
\end{equation}
In fact, besides the non-unitary operator $\hat{U}_a^\dagger$ and unitary operator $\hat{G}_a$, there are infinite different operators which can satisfy $\hat{U}|A\rangle\sim|A\rangle$. For example, let's introduce the Casimir operator for su(1,1) Lie-algebra~\eqref{commtrel},
\begin{equation}\label{Casimir1}
  \hat{C}^{(\vec{k})}:=\hat{L}_0^{(\vec{k})2}-\frac12[\hat{L}_+^{(\vec{k})}\hat{L}_-^{(\vec{k})}+\hat{L}_-^{(\vec{k})}\hat{L}_+^{(\vec{k})}]=\hat{L}_0^{(\vec{k})2}-\hat{L}_0^{(\vec{k})}-\hat{L}_+^{(\vec{k})}\hat{L}_-^{(\vec{k})}\,,
\end{equation}
One can easy check that $[\hat{C}^{(\vec{k})},L^{(\vec{k})}_\pm]=[\hat{C}^{(\vec{k})},L^{(\vec{k})}_0]=0$. Then for any function $h(\vec{k},x)=\sum_{n=0}^{\infty}h_n(\vec{k})x^n$ , the operators,
\begin{equation}\label{Uhak}
\begin{split}
   \hat{O}_1:=&\exp\left[\int\td k^{d-1}e^{-\pi\omega_{\vec{k}}/a}\hat{L}^{(\vec{k})}_++h(\vec{k},\hat{C}^{(\vec{k})})\right]\\
   \hat{O}_2:=&\exp\left[\int\td k^{d-1}e^{-\pi\omega_{\vec{k}}/a}\text{arctanh}e^{-\pi\omega_{\vec{k}}/a}(\hat{L}_+^{(\vec{k})}-\hat{L}_-^{(\vec{k})})+h(\vec{k},\hat{C}^{(\vec{k})})\right]
   \end{split}
\end{equation}
can satisfy the relationship $\hat{O}_1|A\rangle\sim\hat{O}_2|A\rangle\sim|\text{TFD}\rangle$.

\subsection{Complexity between different TFD states}\label{check2bTFD}
In this subsection, let's restrict the generators set into following form,
\begin{equation}\label{TFDgenerE}
  E=\{\hat{L}^{(\vec{k})}_+,\hat{\mathbb{I}}|~\forall \vec{k}\in\mathbb{R}^{d-1}\}\,.
\end{equation}
This generators set contains infinite different generators which are commutative to each others. Similar to what we have argued in the end of subsection~\ref{subcoh}, physical intuition seems to imply that the complexity computed by the generator set \eqref{TFDgenerE} is just the result even when we recover the generator set into the general form given by Eq.~\eqref{extendbasis}. However, the proof is still absent and out of the goal of this paper. Let's restrict the generator set to be Eq.~\eqref{TFDgenerE} in this paper for TFD states.
In the appendix~\ref{app2}, we will use a bigger generator set which contains the Casimir operators $\hat{C}^{(\vec{k})}$ to find the complexity and show the complexity is just the same result given by generator set \eqref{TFDgenerE}. This seems to be evidence for this physical intuition.

Similar to the case in thermal entangle state, any operator $\hat{U}\in\mathcal{U}_E$ has the relationship $\hat{U}\sim\exp(\int\td^{d-1}k\lambda(\vec{k})L^{(\vec{k})}_+)$ for a function $\lambda(\vec{k})\in\mathbb{C}$.  We have to solve the restricted extremum problem,
\begin{equation}\label{Complxity-TFD}
  \mathcal{C}(\hat{U})=\frac{2\text{Vol}}{(2\pi)^{d-1}}\min\left\{\int_0^1\td t\int\td^{d-1}k\parallel y(t,\vec{k})\parallel\right\}
\end{equation}
with the constraint,
\begin{equation}\label{constaintTFD}
\begin{split}
  U\sim\exp\int\td^dk\lambda(\vec{k})\hat{L}^{(\vec{k})}_+=\exp\left\{\hat{L}^{(\vec{k})}_+\int_0^1\td ty(t,\vec{k})\right\}\,.
  \end{split}
\end{equation}
Solving this optimization problem, we can find that
\begin{equation}\label{ComFTDU1}
  \mathcal{C}[\exp\int\td^dk\lambda(\vec{k})\hat{L}^{(\vec{k})}_+]=\frac{2\text{Vol}}{(2\pi)^{d-1}}\int\td^{d-1}k\parallel\lambda(\vec{k})\parallel
\end{equation}
Hence, we can find that,
\begin{equation}\label{complTFD}
  \mathcal{C}(|\text{TFD}\rangle,|A\rangle)=\mathcal{C}(\hat{U}_a^\dagger)=\frac{2\text{Vol}}{(2\pi)^{d-1}}\int\td^{d-1}ke^{-\pi\omega_{\vec{k}}/a}=\frac{2\text{Vol}}{(2\pi)^{d-1}}\int\td^{d-1}ke^{-\omega_{\vec{k}}/(2T)}
\end{equation}
If we assume the quantum field theory has full conformal symmetry, then we have dispersion relationship $\omega_{\vec{k}}=|\vec{k}|$. This leads to following result,
\begin{equation}\label{complTFD}
  \mathcal{C}(|\text{TFD}\rangle,|A\rangle)=\frac{2\text{Vol}}{(2\pi)^{d-1}}\int\td^{d-1}ke^{-k/(2T)}=\frac{2S_{d-2}\Gamma(d-1)}{\pi^{d-1}}\text{Vol}\cdot T^{d-1}\,.
\end{equation}
Here $S_{d-2}$ is the area of $(d-2)$-dimensional unit sphere. It is surprising that the complexity density between the TFD state and its zero temperature vacuum state is finite and proportional to $T^{d-1}$. This is just the behavior of renormalized holographic complexity in Schwarzschild-AdS black hole with planar symmetry~\cite{Kim:2017lrw}!  In addition, the result~\eqref{complTFD} seems to be against the expectations in Refs.~\cite{Lehner:2016vdi,Jefferson:2017sdb,Carmi:2016wjl,Chapman:2017rqy,Caputa:2017yrh} that the complexity density about a TFD state should be infinite. To clarify why the complexity density between a TFD state and its corresponding vacuum state should be finite, I will make some detailed discussions in Sec.~\ref{whyfinite}.

Beside the decomposition $\mathcal{H}=\mathcal{H}_R\times\mathcal{H}_L$ and $\mathcal{H}=\mathcal{H}_U\times\mathcal{H}_D$, we can also make a new decomposition $\mathcal{H}=\mathcal{H}_B\times\mathcal{H}_W$ and its creation/annihilation operators group $\{\hat{c}^{B\dagger}_{\vec{k}_i}, \hat{c}^{B}_{\vec{k}_i}, \hat{c}^{W\dagger}_{\vec{k}_i},\hat{c}^{W}_{\vec{k}_i}\}$. Then these annihilation operators define a vacuum $|C\rangle\in\mathcal{H}$ such that $\hat{c}^{W}_{\vec{k}_i}|C\rangle=\hat{c}^{D}_{\vec{k}_i}|C\rangle=0$. 
Then we define Bogoliubov transformations between them as follows,
\begin{equation}\label{bogoliu2}
  \hat{b}_{\vec{k}}^{U}\propto\hat{a}_{\vec{k}}^{R}-e^{-\pi\omega_{\vec{k}}/a_1}\hat{a}_{\vec{k}}^{L\dagger}, ~~\hat{b}_{\vec{k}}^{D}\propto\hat{a}_{\vec{k}}^{L}-e^{-\pi\omega_{\vec{k}}/a_1}\hat{a}_{\vec{k}}^{R\dagger}\,.
\end{equation}
and,
\begin{equation}\label{bogoliu3}
  \hat{c}_{\vec{k}}^{W}\propto\hat{a}_{\vec{k}}^{R}-e^{-\pi\omega_{\vec{k}}/a_2}\hat{a}_{\vec{k}}^{L\dagger}, ~~\hat{c}_{\vec{k}}^{B}\propto\hat{a}_{\vec{k}}^{L}-e^{-\pi\omega_{\vec{k}}/a_2}\hat{a}_{\vec{k}}^{R\dagger}\,.
\end{equation}
They can give two TFD states,
\begin{equation}\label{TFDdep3a}
\begin{split}
  |\text{TFD}_1\rangle\sim&\exp\left[\int\td k^{d-1}e^{-\pi\omega_{\vec{k}}/a_1}\hat{L}^{(\vec{k})}_+\right]|A\rangle\\
|\text{TFD}_2\rangle\sim&\exp\left[\int\td k^{d-1}e^{-\pi\omega_{\vec{k}}/a_2}\hat{L}^{(\vec{k})}_+\right]|A\rangle
  \end{split}
\end{equation}
We can find that,
\begin{equation}\label{complTFD1}
  \mathcal{C}(|\text{TFD}_1\rangle,|A\rangle)=\frac{2\text{Vol}}{(2\pi)^{d-1}}\int\td^{d-1}ke^{-\omega_{\vec{k}}/(2T_1)}\,.
\end{equation}
and,
\begin{equation}\label{complTFD2}
  \mathcal{C}(|\text{TFD}_2\rangle,|A\rangle)=\frac{2\text{Vol}}{(2\pi)^{d-1}}\int\td^{d-1}ke^{-\omega_{\vec{k}}/(2T_2)}\,.
\end{equation}
Here $T_1=a_1/(2\pi)$ and $T_2=a_2/(2\pi)$. We see that the complexity of formation between this two TFD states with respective to reference vacuum state $|A\rangle$,
\begin{equation}\label{complfor12}
  \Delta\mathcal{C}_A(|\text{TFD}_2\rangle,|\text{TFD}_1\rangle)=\frac{2\text{Vol}}{(2\pi)^{d-1}}\int\td^{d-1}k\left[e^{-\omega_{\vec{k}}/(2T_2)}-e^{-\omega_{\vec{k}}/(2T_1)}\right]\,.
\end{equation}

As the operators $\hat{a}^{R\dagger}(\vec{k}_1)\hat{a}^{L\dagger}(\vec{k}_1)$ and $\hat{a}^{R\dagger}(\vec{k}_2)\hat{a}^{L\dagger}(\vec{k}_2)$ are commutated with each other for any two momentum $\vec{k}_1$ and $\vec{k}_2$, the complexity between any to TFD states will be given by the manner similar to the subsection~\ref{ETS1}, which reads,
\begin{equation}\label{cmplexU}
\begin{split}
  \mathcal{C}(|\text{TFD}_1\rangle,|\text{TFD}_2\rangle)&=\frac{2\text{Vol}}{(2\pi)^{d-1}}\int\td^{d-1}k|e^{-\omega_{\vec{k}}/(2T_2)}-e^{-\omega_{\vec{k}}/(2T_1)}|\,.
  \end{split}
\end{equation}
We see that the absolute value of complexity of formation according to the vacuum state $|A\rangle$ is just the complexity between them. In fact, one can show that, by choosing any $|\text{TFD}_3\rangle$ as the reference state, the complexity of formation is given by,
\begin{equation}\label{complfor2}
\begin{split}
  &\Delta\mathcal{C}_{\text{TFD}_3}(|\text{TFD}_2\rangle,|\text{TFD}_1\rangle)\\
  =&\frac{2\text{Vol}}{(2\pi)^{d-1}}\int\td^{d-1}k|e^{-\omega_{\vec{k}}/(2T_2)}-e^{-\omega_{\vec{k}}/(2T_3)}| -|e^{-\omega_{\vec{k}}/(2T_3)}-e^{-\omega_{\vec{k}}/(2T_1)}|\,.
  \end{split}
\end{equation}
Hence, just as the same as the case in (1+0)-dimensional TFD stats shown in the subsection~\ref{ETS1}, if the temperatures of TFD states satisfy that $T_3\leq\min\{T_1,T_2\}$ or $T_3\geq\max\{T_1,T_2\}$, then the complexity of formation is independent of the choice on reference TFD state and its absolute value is just the complexity between two states, i.e.,
\begin{equation}\label{complfor3}
\begin{split}
  &|\Delta\mathcal{C}_{\text{TFD}_3}(|\text{TFD}_2\rangle,|\text{TFD}_1\rangle)|=\mathcal{C}(|\text{TFD}_2\rangle,|\text{TFD}_1\rangle),\\
  \text{if}~&T_3\leq\min\{T_1,T_2\}~\text{or}~T_3\geq\max\{T_1,T_2\}\,.
  \end{split}
\end{equation}
The physical reason for that is just as the same as what was shown in the subsection~\ref{ETS1}.

Specially, for the case that the system has full conformal symmetry, we have that,
\begin{equation}\label{complfor2b}
\begin{split}
  &\Delta\mathcal{C}_{\text{TFD}_3}(|\text{TFD}_2\rangle,|\text{TFD}_1\rangle)\\
  =&\frac{2S_{d-2}\Gamma(d-1)}{\pi^{d-1}}\text{Vol}\cdot (|T_2^{d-1}-T_3^{d-1}|-|T_1^{d-1}-T_3^{d-1}|)\,.
  \end{split}
\end{equation}
If $T_3\leq\min\{T_1,T_2\}$ or $T_3\geq\max\{T_1,T_2\}$, Eq.~\eqref{complfor2b} then becomes,
\begin{equation}\label{complfor4}
\begin{split}
  \mathcal{C}_{\text{TFD}_3}(|\text{TFD}_2\rangle,|\text{TFD}_1\rangle)&=|\Delta\mathcal{C}_{\text{TFD}_3}(|\text{TFD}_2\rangle,|\text{TFD}_1\rangle)|\\
  &=\frac{2S_{d-2}\Gamma(d-1)}{\pi^{d-1}}\text{Vol}\cdot|T_2^{d-1}-T_1^{d-1}|\,.
  \end{split}
\end{equation}
This result shows that in these case the complexity of two TFD states is just the complexity of formation.

It needs to emphasis that if $T_3$ is between $T_2$ and $T_1$, then the complexity of formation $\Delta\mathcal{C}_{\text{TFD}_3}(|\text{TFD}_2\rangle,|\text{TFD}_1\rangle)$ will not be equivalent to the complexity between $|\text{TFD}_2\rangle$ and $|\text{TFD}_1\rangle$. In this case, $\Delta\mathcal{C}_{\text{TFD}_3}(|\text{TFD}_2\rangle,|\text{TFD}_1\rangle)$ will depends on the value of $T_3$. For example, let's assume $T_1\leq T_3\leq T_2$, then one can find that,
\begin{equation}\label{complfor5}
  \Delta\mathcal{C}_{\text{TFD}_3}(|\text{TFD}_2\rangle,|\text{TFD}_1\rangle)=\frac{2S_{d-2}\Gamma(d-1)}{\pi^{d-1}}\text{Vol}\cdot(T_2^{d-1}+T_1^{d-1}-2T_3^{d-1})\,.
\end{equation}
Obviously, this result depends on the value of $T_3$ and can satisfy the inequality~\eqref{formC2}.

\subsection{Equivalence to fidelity susceptibility}\label{Fideli2C}
Now I will try to connect the other useful conception in the quantum information theory, the fidelity susceptibility (or information metric), to the conception of complexity. More precisely, I will show that the fidelity susceptibility for a TFD state in fact is equivalent to the complexity between it and corresponding vacuum state. This statement is motivated by Ref.~\cite{MIyaji:2015mia}, which gives a proposal that the fidelity susceptibility  of a TFD state is given by the maximum volume of space-like surfaces which connect the two boundary of an enteral asymptotic AdS black hole. We see that the holographic objects of fidelity susceptibility  and complexity in CV conjecture are the same one. This gives us strong evidence and motivation to connect two conceptions.

Let's assume that $|\psi(\lambda)\rangle$ to be a curve in Hilbert space $\mathcal{H}$ and $|\psi(0)\rangle=|\psi_0\rangle$. In general, this curve can be generated by a $\lambda$-dependent tangent operator $\hat{T}(\lambda)$, i.e.,
\begin{equation}\label{generUF}
  |\psi(\lambda)\rangle=\frac1{\mathcal{N}(\lambda)}\overleftarrow{\mathcal{P}}\exp\left[\int_0^\lambda\hat{T}(s)\td s\right]|\psi_0\rangle\,.
\end{equation}
Here $\mathcal{N}(\lambda)\in\mathbb{R}^+$ is the normalization factor so that $\langle\psi(\lambda)|\psi(\lambda)\rangle=1$. If $\hat{T}(\lambda)$ is anit-Hermit, i.e., $\hat{T}^\dagger(\lambda)=-\hat{T}(\lambda)$, then $\mathcal{N}(\lambda)=1$. 
In general case, we have $\mathcal{N}(\lambda)\neq1$. Then the projection of $|\psi(\lambda)\rangle$ on $|\psi_0\rangle$ is given by $\langle\psi_0|\psi(\lambda)\rangle$. The fidelity susceptibility $G_{T}$ (or information metric) then is given by,
\begin{equation}\label{defGTT}
  G_{\hat{T}}:=\lim_{\lambda\rightarrow0}\frac1{\lambda^2}[1-|\langle\psi_0|\psi(\lambda)\rangle|]=\lim_{\lambda\rightarrow0}\frac1{\lambda^2}\left\{1-\frac1{\mathcal{N}(\lambda)}\left|\langle\psi_0|\exp\left[\int_0^\lambda\hat{T}(s)\td s\right]|\psi_0\rangle\right|\right\}\,.
\end{equation}
We see that fidelity susceptibility depends on the state $|\psi_0\rangle$ and generator $\hat{T}_0:=\hat{T}(\lambda)|_{\lambda=0}$. We can write Eq.~\eqref{defGTT} into a more explicate form. One can easy check that,
\begin{equation}\label{defGTT2}
  2G_{\hat{T}}=\langle\psi_0|\hat{T}_0\hat{T}^\dagger_0|\psi_0\rangle-\langle\psi_0|\hat{T}_0|\psi_0\rangle\langle\psi_0|\hat{T}^\dagger_0|\psi_0\rangle\,.
\end{equation}
If we look at the proposal about the definition of complexity according to Fubini-Study metric in Ref.~\cite{Chapman:2017rqy}, then we see that Eq.~\eqref{defGTT2} is nothing but a line element in Fubini-Study metric. If we use the method in  Ref.~\cite{Chapman:2017rqy} to define the complexity, then the fidelity susceptibility measures the ``infinitesimal complexity'' for nearby two states.

For the definition of the complexity in this paper, it is not easy to find its relationship to fidelity susceptibility. Let's first consider an explicit example by computing  fidelity susceptibility of a TFD state. For a given theory, there is only one parameter, the temperature $T$, to describe different TFD states\footnote{The time evolution of TFD states are not considered here}. Assume the $|\psi_\beta\rangle$ is a TFD state with temperature $T=1/\beta$, which is defined according to a vacuum $|A\rangle$ state in this way,
\begin{equation}\label{defTFDt1}
  |\psi_\beta\rangle:=\frac1{\mathcal{N}}\exp\left[\int\td k^{d-1}e^{-\beta\omega_{\vec{k}}/2}\hat{L}^{(\vec{k})}_+\right]|A\rangle
\end{equation}
Here the factor $\mathcal{N}$ is applied so that $|\psi_\beta\rangle$ is normalized. Now consider one parameter family of TFD states $|\psi_\beta(\lambda)\rangle:=|\psi_{\beta(1+\lambda)}\rangle$, which is generated by,
\begin{equation}\label{defTFDt1}
\begin{split}
  |\psi_\beta(\lambda)\rangle:&=\frac1{\mathcal{N}(\lambda)}\exp\left[\int\td k^{d-1}(e^{-\lambda\beta\omega_{\vec{k}}/2}-1)e^{-\beta\omega_{\vec{k}}/2}\hat{L}^{(\vec{k})}_+\right]|\psi_\beta\rangle\\
  &=\frac1{\mathcal{N}(\lambda)}\exp\left[-\int_0^\lambda \td s\int\td k^{d-1} \frac{\beta\omega_{\vec{k}}}2e^{-(1+s)\beta\omega_{\vec{k}}/2}\hat{L}^{(\vec{k})}_+\right]|\psi_\beta\rangle\,.
  \end{split}
\end{equation}
Then we can read that,
\begin{equation}\label{generaTl}
  \hat{T}_0=-\frac{\beta}2\int\td k^{d-1} \omega_{\vec{k}}e^{-\beta\omega_{\vec{k}}/2}\hat{L}^{(\vec{k})}_+\,.
\end{equation}
Using Eq.~\eqref{defGTT2}, we can reads,
\begin{equation}\label{defGTT3}
\begin{split}
  2G_{\hat{T}}=&\langle \psi_\beta|\hat{T}^\dagger_0\hat{T}_0|\psi_\beta\rangle-\langle \psi_\beta|\hat{T}^\dagger_0|\psi_\beta\rangle\langle\psi_\beta|\hat{T}_0|\psi_\beta\rangle\\
  &=\frac{\beta^2}4\int\td k^{d-1} \omega_{\vec{k}}^2e^{-\beta\omega_{\vec{k}}}\langle \psi_\beta|\hat{L}^{(\vec{k})}_-\hat{L}^{(\vec{k})}_+|\psi_\beta\rangle\,.
  \end{split}
\end{equation}
This equation is not easy to computed in general if we directly take Eqs.~\eqref{defTFDt1} and \eqref{generaTl} into Eq.~\eqref{defGTT3}. However,  as what we have shown in the section~, the TFD state is the vacuum state of annihilation operators $\{\hat{b}^{U\dagger}(\vec{k}), \hat{b}^{D\dagger}(\vec{k})\}$ defined by Eq.~\eqref{bogoliu1}, i.e., $\hat{b}^{U}(\vec{k})|\psi_\beta\rangle=\hat{b}^{D}(\vec{k})|\psi_\beta\rangle=0$  and $\beta=2\pi/a$. then we can expressed $\hat{L}^{(\vec{k})}_+$ by,
\begin{equation}\label{a2b1}
\begin{split}
  \hat{L}^{(\vec{k})}_+&=\hat{a}^{R\dagger}(\vec{k}) \hat{a}^{L\dagger}(\vec{k})=\sinh^2\xi_{\vec{k}}\hat{b}^{U}(\vec{k})\hat{b}^{D}(\vec{k})+\cosh^2\xi_{\vec{k}}\hat{b}^{U\dagger}(\vec{k})\hat{b}^{D\dagger}(\vec{k})\\ &-\frac{\sinh2\xi_{\vec{k}}}2[\hat{b}^{D}(\vec{k})\hat{b}^{D\dagger}(\vec{k})+\hat{b}^{U\dagger}(\vec{k})\hat{b}^{U}(\vec{k})]\,.
  \end{split}
\end{equation}
Here $\xi_{\vec{k}}:=e^{-\beta\omega_{\vec{k}}/2}$. After some algebras, we can find that
\begin{equation}\label{meanT1}
  \langle\psi_\beta|\hat{L}^{(\vec{k})}_-\hat{L}^{(\vec{k})}_+|\psi_\beta\rangle=\frac1{4}\frac{\text{Vol}}{(2\pi)^{d-1}}(\cosh4\xi_{\vec{k}}-2\cosh2\xi_{\vec{k}}+1)\,,
\end{equation}
We finally find  that the fidelity susceptibility of a TFD state is
\begin{equation}\label{defGTT4}
\begin{split}
  G_{\hat{T}}=\frac{\beta^2}{32}\frac{\text{Vol}}{(2\pi)^{d-1}}\int\td k^{d-1} \omega_{\vec{k}}^2e^{-\beta\omega_{\vec{k}}}(\cosh4\xi_{\vec{k}}-2\cosh2\xi_{\vec{k}}+1)\,.
  \end{split}
\end{equation}
In full conformal symmetry case, we have $\omega_{\vec{k}}=k$. Thus Eq.~\eqref{defGTT4} reads
\begin{equation}\label{defGTT4b}
\begin{split}
  G_{\hat{T}}=\frac{\text{Vol}~S_{d-2}\vartheta_d}{32(2\pi)^{d-1}}\beta^{1-d}\,.
  \end{split}
\end{equation}
with,
\begin{equation}\label{defvarnu0}
  \vartheta_d=\int_0^\infty x^de^{-x}(\cosh4e^{-x}-2\cosh2e^{-x}+1)\td x\,.
\end{equation}
On the other hand, we can see from Eq.~\eqref{complTFD} that the complexity between $|\psi_\beta\rangle$ and $|A\rangle$ reads
\begin{equation}\label{complld}
  \mathcal{C}(|\psi_\beta\rangle,|A\rangle)=\frac{2S_{d-2}\Gamma(d)}{\pi^{d-1}}\text{Vol}\cdot\beta^{1-d}\,.
\end{equation}
Combining Eqs.~\eqref{defGTT4} and \eqref{complld}, we see that,
\begin{equation}\label{complld}
  \mathcal{C}(|\psi_\beta\rangle,|A\rangle)=\frac{2^{d+5}\Gamma(d)}{\vartheta_d}G_{\hat{T}}\,.
\end{equation}
This equation clearly shows that the fidelity susceptibility is equivalent to the complexity between a TDF state and corresponding vacuum state. This result seems to supply an explanation on why two quantities may share the same holographic object.
This subsection only shows that the fidelity susceptibility and complexity are equivalent to each for TDF states. It is not clear if such  equivalence can happen in more general states.

\section{Discussion}\label{whyfinite}

\subsection{Reasons of finite complexity density}
In the subsection~\ref{TFD2}, it has been shown that the complexity desnity between a TFD state and the zero temperature vacuum state is finite. This seems to be against with the  results from the holographical duality such as Refs.~\cite{Lehner:2016vdi,Carmi:2016wjl,Caputa:2017yrh} and some anther attempts for building complexity from field theory frameworks such as Refs.~\cite{Jefferson:2017sdb,Chapman:2017rqy}. In following, I will explain that this is because of the different choices on reference states or systems.

The complexity here is defined between two states rather than for one state. When someone tries to ask what is the complexity for one state, he must first clarify the reference state in a very clear manner. The indistinct announce such as ``choosing a reference state'' is a little ambiguous since there is not a unique quantum state in Hilbert space which is simpler than all other states. Even a vacuum state is also a kind of TFD state for some particular choice on annihilation operators. In the subsection~\ref{TFD2}, for a TFD state, the reference state $|A\rangle$ is chosen so that it satisfies the Eq.~\eqref{TFDdep3}. Then one can see that in the UV region $|\vec{k}|\rightarrow\infty$, the coefficient $e^{-\pi\omega_{\vec{k}}/a}\rightarrow0$ and the TFD state in fact just inherits the UV structure of reference state. This implies that we need not to add quantum gates to change the UV part and the UV divergence of gate number will not appear.
\begin{figure}
  \centering
  \includegraphics[width=.9\textwidth]{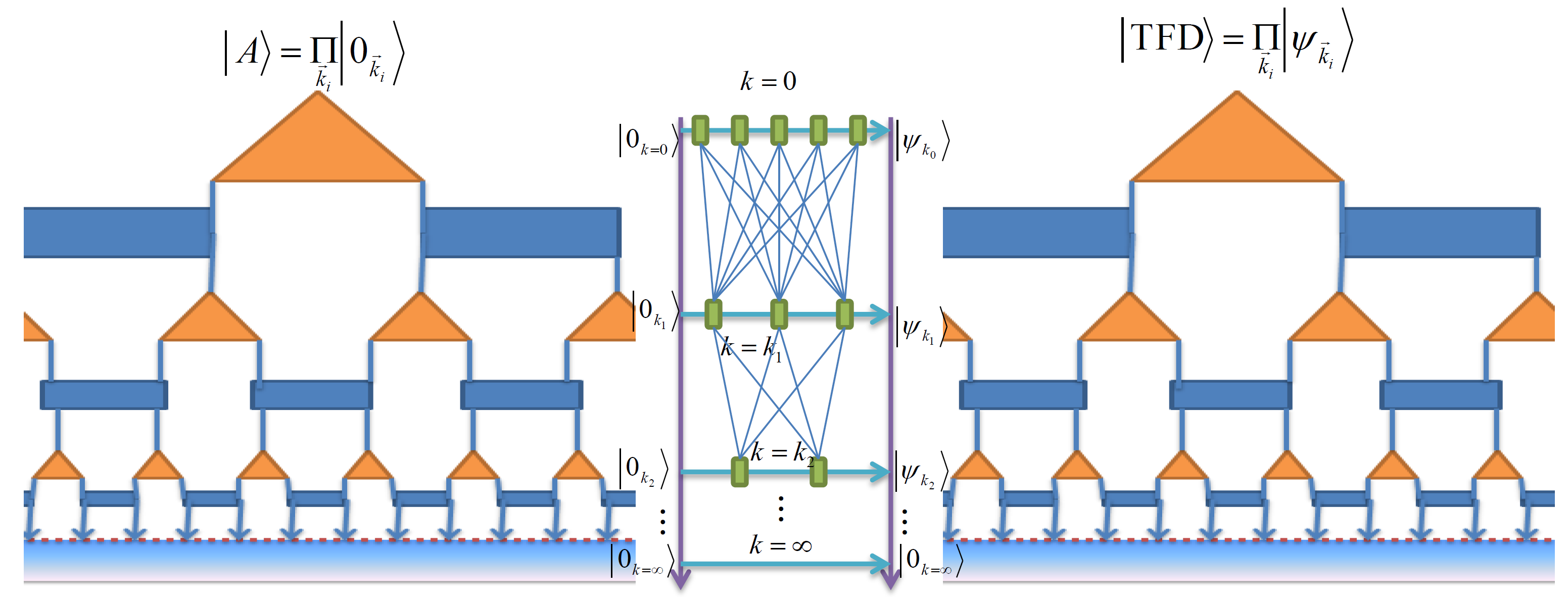}
  \caption{The schematic explanation about why the complexity from $|A\rangle$ to a TFD state is finite. The left and right sides are MERA approximations for states $|A\rangle$ and $|\text{TFD}\rangle$ in tensor network representation. The some green rectangles at the middle of two tensor networks stand for the quantum circuit. The horizonal direction stands for the $(d-1)$-dimensional spatial directions and the vertical direction is the length scale (inverse of momentum).  For convenience, the spatial direction and momentum are shown in 1-dimensional case and only one copy of a double state is shown in the figure. One can image that the $i$-th layer is the state $|\psi_{k_i}\rangle:=\sum_{n=0}^\infty e^{-\pi n\omega_{\vec{k}_i}/a}|n_{i}\rangle_L |n_{i}\rangle_R$.  }\label{TNRG1}
\end{figure}
In Refs.~\cite{Lehner:2016vdi,Carmi:2016wjl,Caputa:2017yrh}, the reference state for computing the complexity of a TFD state are chosen as a kind of particular ``simple'' reference state. Though the physical properties of this reference state are not clear, it is not the zero temperature vacuum state corresponding to this TFD state as the complexity between this reference state and the vacuum state is not zero. Hence, the infinity discussed by Refs.~\cite{Lehner:2016vdi,Carmi:2016wjl,Caputa:2017yrh} and the finite shown in this paper are not contradictory as the reference states are different. Refs.~\cite{Jefferson:2017sdb,Chapman:2017rqy} studied the complexity for some operators and between the states which are different from the TFD states, so the results are different. It will be shown later that, whence the UV structures of two states are different, the complexity between them is divergent.

To understand more clearly about why the complexity between a TDF state and its corresponding zero temperature vacuum state is finite, let's consider  MERA approximations for states $|A\rangle$ and $|\text{TFD}\rangle$ in tensor network representation. In the Fig.~\ref{TNRG1}, the MERA approximations for $|A\rangle$ and $|\text{TFD}\rangle$ are shown. The left and right sides are MERA approximations for states $|A\rangle$ and $|\text{TFD}\rangle$ in tensor network representation. The horizonal direction stands for the $(d-1)$-dimensional spatial directions and the vertical direction is the length scale (inverse of momentum).  For convenience, the spatial direction and momentum are shown in 1-dimensional case. The green rectangles at the middle of two tensor networks stand for the quantum circuit. In the Fig.~\ref{TNRG1}, only one copy of a double state is show, i.e., only the parts belong to $\mathcal{H}_L$ are shown. In principle, one should add the other copy at the head of Fig.~\ref{TNRG1} to stand for the parts belong to $\mathcal{H}_R$. However, the simplified schematic figure is enough for our purpose.

The disentanglers (blue rectangles) and isometries (triangles) in Fig.~\ref{TNRG1} connect the microscopic degrees of freedom at the very layer (length scale). In the full MERA approximation, the tensors of $|A\rangle$ and $|\text{TFD}\rangle$  are both infinite. This divergence is the divergent discussed in previous papers such as Refs.~\cite{Lehner:2016vdi,Jefferson:2017sdb,Carmi:2016wjl,Chapman:2017rqy,Caputa:2017yrh}. However, the complexity in this paper is not defined by how many tensors are needed when we use  MERA to approximate a particular state. We define the complexity is the minimal required logic gates when we use a quantum circuit to convert one state to the other, which can be presented by the green rectangles in the middle region of two tensor networks in Fig.~\ref{TNRG1}. With increasing the momentum, the disentangle and isometry tensors are increased in order of $|\vec{k}|^{d-1}$. However, the Bogoliubov transformation~\eqref{bogoliu1} shows that average  particles number that we have to add into vacuum state decays exponentially in the order of $e^{-\pi\omega_{\vec{k}}/a}$. This means that the deeper layer of TFD state will inherit the more UV tensors and the change compared with the corresponding parts in $|A\rangle$ is suppressed  exponentially. Hence, the number of gates will decease in UV region when we increase the circuit depth, though the tensors in MERA for both $|A\rangle$ and $|\text{TFD}\rangle$ are increased. At the UV limit, the the disentangle and isometry tensors are divergent but the gates number in the quantum circuit is zero. As a result, the total gates in the Fig.~\ref{TNRG1} is finite. By this explanation, we see that the finite complexity between a TFD state and its corresponding zero temperature limit vacuum state is just the result of that they share the same UV structure. Based on a similar arguments, one can also see that the complexity between two different TFD states defined by Eq.~\eqref{TFDdep3a} is also finite.

\subsection{Enlightenments to holographic conjectures}
It needs to note that computations in Secs.~\ref{sim-exmp} and \ref{Com-TFD} do not involve the dynamics of the fields.  In order to make a connection to the results in Refs.~\cite{Lehner:2016vdi,Jefferson:2017sdb,Carmi:2016wjl,Chapman:2017rqy,Caputa:2017yrh}, the detailed model and the dynamics are needed. Let's assume the quantum field theory is a free scalar field with Hamilton,
\begin{equation}\label{scalarL1}
  \hat{H}=\frac12\int\td^{d-1}x\left[\hat{\pi}^2+(\vec{\nabla}\hat{\phi})^2+m^2\hat{\phi}^2\right]\,,
\end{equation}
and the minimal energy state corresponding to this Hamilton is the vacuum state $|A\rangle$. This means that the creation and annihilation operators $\{\hat{a}^{R}(\vec{k}), \hat{a}^{L}(\vec{k}),\hat{a}^{R\dagger}(\vec{k}), \hat{a}^{L\dagger}(\vec{k})\}$ have following relationship to the scalar field operator in momentum space,
\begin{equation}\label{scalars1}
  \hat{\phi}(\vec{k})=\hat{\phi}^L(\vec{k})+\hat{\phi}^R(\vec{k})
\end{equation}
with
\begin{equation}\label{scalars2}
  \hat{\phi}^L(\vec{k})=\frac1{\sqrt{2\omega_{\vec{k}}}}[\hat{a}^{L}(\vec{k})+\hat{a}^{L\dagger}(-\vec{k})],~~\hat{\phi}^R(\vec{k})=\frac1{\sqrt{2\omega_{\vec{k}}}}[\hat{a}^{R}(\vec{k})+\hat{a}^{R\dagger}(-\vec{k})]\,.
\end{equation}
One can see that state $|A\rangle$ minimizes the expected value of Hamilton \eqref{scalarL1}, i.e., $\langle A|:\hat{H}:|A\rangle=0$. Let's choose special state $|Q(\xi)\rangle$ as the reference state rather than the $|A\rangle$ or TFD states, which is the vacuum state corresponding to annihilation operators $\hat{q}^L(\vec{k})$ and $\hat{q}^R(\vec{k})$. The relationships between $\{\hat{q}^{R}(\vec{k}), \hat{q}^{L}(\vec{k}), \hat{q}^{L\dagger}(\vec{k}),\hat{q}^{R\dagger}(\vec{k})\}$ and $\{\hat{a}^{R}(\vec{k}), \hat{a}^{L}(\vec{k}),\hat{a}^{R\dagger}(\vec{k}), \hat{a}^{L\dagger}(\vec{k})\}$ are given by,
\begin{equation}\label{aandq1}
  \left[\begin{matrix}
  \hat{q}^L(\vec{k})\\
  \hat{q}^{R\dagger}(\vec{k})
  \end{matrix}
  \right]=\left[\begin{matrix}
  \cosh\xi_{\vec{k}}&-\sinh\xi_{\vec{k}}\\
  -\sinh\xi_{\vec{k}}&\cosh\xi_{\vec{k}}
  \end{matrix}\right]\left[\begin{matrix}
  \hat{a}^L(\vec{k})\\
  \hat{a}^{R\dagger}(\vec{k})
  \end{matrix}\right]
\end{equation}
with the parameter $\xi_{\vec{k}}$. If one take $\tanh\xi_{\vec{k}}=e^{-\pi\omega_{\vec{k}}/a}$, then the state $|Q(\xi)\rangle$ is just the TFD state shown in Eq.~\eqref{TFDdep3}. It can be proven that
 $$\langle Q(\xi)|\hat{\phi}(\vec{k})\hat{\phi}(\vec{k}')|Q(\xi)\rangle=\frac{e^{-2\xi_{\vec{k}}}}{2\omega_{\vec{k}}}\delta^{d-1}(\vec{k}+\vec{k}')$$
Converting it into the spatial coordinate, we can read that,
\begin{equation}\label{correl1}
  \langle Q(\xi)|\hat{\phi}(\vec{x})\hat{\phi}(\vec{x}')|Q(\xi)\rangle=\int\td^{d-1}k\frac{e^{-2\xi_{\vec{k}}}}{2\omega_{\vec{k}}}e^{i\vec{k}\cdot(\vec{x}-\vec{x}')}
\end{equation}

By similar steps in subsection~\ref{TFD2}, one can find the following relationship between $|A\rangle$ and $|Q(\xi)\rangle$,
\begin{equation}\label{relAQ}
  |Q(\xi)\rangle\sim\exp\left[\int\td^{d-1}k\tanh\xi_{\vec{k}}\hat{a}^{R\dagger}(\vec{k})\hat{a}^{L\dagger}(\vec{k})\right]|A\rangle
\end{equation}
Let's assume $|\text{TFD}_a\rangle:=\hat{U}^\dagger_a|A\rangle$ where $\hat{U}^\dagger_a$ is given by \eqref{defUa}. Then it is easy to find that $|A\rangle\sim\exp\left[-\tanh\xi_{\vec{k}}\int\td k^{d-1}\hat{a}^{R\dagger}(\vec{k})\hat{a}^{L\dagger}(\vec{k})\right]|Q(\xi)\rangle$ and,
\begin{equation}\label{TFDdep4}
  |\text{TFD}_a\rangle\sim\exp\left[\int\td k^{d-1}(e^{-\pi\omega_{\vec{k}_i}/a}-\tanh\xi_{\vec{k}})\hat{a}^{R\dagger}(\vec{k})\hat{a}^{L\dagger}(\vec{k})\right]|Q(\xi)\rangle\,
\end{equation}
Under the reduced generator set in Eq.~\eqref{TFDgenerE}, we see that,
\begin{equation}\label{complTFD3}
  \frac{(2\pi)^{d-1}}{2\text{Vol}}\mathcal{C}(|\text{TFD}_a\rangle,|Q(\xi)\rangle)=\int\td^{d-1}k|\tanh\xi_{\vec{k}}-e^{-\pi\omega_{\vec{k}}/a}|\,.
\end{equation}
If one take the parameter,
\begin{equation}\label{valuegamma}
  \xi_{\vec{k}}=\frac12\ln(M/\omega_{\vec{k}})\,
\end{equation}
for arbitrary energy scala $M$,  then Eq.~\eqref{correl1} becomes,
\begin{equation}\label{correl2}
  \langle Q(\xi)|\hat{\phi}(\vec{x})\hat{\phi}(\vec{x}')|Q(\xi)\rangle=\frac1{2M}\delta^{d-1}(\vec{x}-\vec{x}')\,.
\end{equation}
In this case, the state $|Q(\xi)\rangle$ is an unentangled product state, which has no any spatial correlations.  This state appears in Refs.~\cite{Haegeman:2011uy,Hu:2017rsp} as initial product state to construct the vacuum state $|A\rangle$ in cMERA (the continuous version of MERA) and is the reference state in Ref.~\cite{Chapman:2017rqy} to compute the complexity by Fubini-Study metric. Taking the Eq.~\eqref{valuegamma} into the expression~\eqref{complTFD3}, one can see that the integration is divergent when $|\vec{k}|\rightarrow\infty$. This result explicitly shows that the complexity between a TFD state and a particular reference state may be infinite. What's more, after a UV cut-off $k_m=M=1/\delta$ is introduced with a small length scala $\delta$ and the full conformal symmetry is imposed, we can see that the divergence in Eq.~\eqref{complTFD3} is,
\begin{equation}\label{complTFD4}
  \frac{(2\pi)^{d-1}}{2\text{Vol}}\mathcal{C}(|Q(\xi)\rangle,|\text{TFD}_a\rangle)\propto\frac1{\delta^{d-1}}+\text{finite term}\,.
\end{equation}
The divergent structure is the same as the one studied by Refs.~\cite{Carmi:2016wjl,Reynolds:2016rvl,Caputa:2017yrh} and also appears in Ref.~\cite{Chapman:2017rqy}.\footnote{In fact, the subleading divergent terms can appear in both CV and CA conjectures if the time slices at the boundary are not flat. This corresponds the deformed conformal field theory rather than a free conformal field theory discussed in this subsection. } However, the choice in Eq.~\eqref{valuegamma} is not the unique reference state to match the divergent structures of CV or CA conjecture. One can just take $\xi_{\vec{k}}$ to be any nonzero constant, then he can still obtain the result shown in Eq.~\eqref{complTFD4}. Thus, we see that there are infinite different states, which can be the reference states and give the divergent structure just as the same as ones in CV or CA conjectures. This seems to imply that, it seems hard to clarify what is the reference state in CV and CA conjectures just by studying the divergent structure of complexity. However, because of the results about the complexity between two TFD states in subsection~\ref{check2bTFD}, it is still possible that both these two conjectures do not give the complexity for the TFD state but the absolute value of the difference in two black holes gives the complexity between corresponding two TFD states. In this modified version, the reference state is not needed. Both CV and CA conjecture in fact give the some kind of ``complexity potential''. In order to compute the complexity between two TFD states $|\text{TFD}_1\rangle$ and $|\text{TFD}_2\rangle$, we need to use Eqs.~\eqref{CV} or \eqref{CA} to compute the corresponding $\mathcal{C}^{(1)}_{V}$ and $\mathcal{C}^{(2)}_{V}$ or $\mathcal{C}^{(1)}_{A}$ and $\mathcal{C}^{(2)}_{A}$, then the complexity between $|\text{TFD}_1\rangle$ and $|\text{TFD}_2\rangle$ is given by
\begin{equation}\label{newcvca}
  \mathcal{C}_V(|\text{TFD}_2\rangle,|\text{TFD}_1\rangle)=|\mathcal{C}^{(1)}_{V}-\mathcal{C}^{(2)}_{V}|,~~\text{or}~\mathcal{C}_A(|\text{TFD}_2\rangle,|\text{TFD}_1\rangle)=|\mathcal{C}^{(1)}_{A}-\mathcal{C}^{(2)}_{A}|\,.
\end{equation}
In fact, this modified holographic version does not lose important physical properties of the original version and seems to be simpler as it does not need to refer to an unknown reference state. In addition, it just matches the results obtained in field theory approach in Sec.~\ref{check2bTFD}. It is interesting and worthy of investigating this idea further.

\section{Summary}\label{summ}
Let's make a brief summary. By this paper, the complexity between two states in quantum field theory was studied by introducing a Finsler structure based on ladder operators. Some simple examples, including coherent states and entangled thermal states, were computed to show how to use this method and clarify the differences between complexity and other conceptions such as complexity of formation and entanglement entropy. Then this method was applied to compute complexity of two thermofield double states.  The results showed that the complexity density between a thermofield double state and corresponding zero temperature ground state was finite. In addition,  it was found that complexity for $d$-dimension conformal field showed the behavior of $\mathcal{C}\propto T^{d-1}$, which is just the the renormalized complexity predicted by CA and CV conjectures. It has also been shown that fidelity susceptibility of a TFD state is equivalent to the complexity between it and corresponding vacuum state, which gave an explanation why they could share the same object in holographic duality. It was also showed that if the reference state and TFD state had different UV structures, the complexity between them was divergent. Especially, for some reference state, the method in this paper gave the same divergent structure in the CV and CA conjectures. The results in this paper seem to imply that the difference of volumes or actions in two black holes computed by CV or CA conjectures might correspond to the complexity between two TFD states.

Though the computations for TFD states were done in scalar field, it is no any essential difficulty to generalize them into higher spin bosonic fields and obtain some similar results. For fermi fields, the Bogoliubov transformations from vacuum state to TFD state are different from the forms in Eq.~\eqref{bogoliu1}. This leads to some important differences which are worthy of investigating in future. It is also very interesting to use this method to study the growth rate of time-dependent TFD state defined in Eq.~\eqref{timesate1}. Especially, the CV and CA conjectures give different predictions for complexity growth rate at the early time. The CV conjecture shows that the complexity growth rate is finite at early time~\cite{Hartman:2013qma,MIyaji:2015mia}. However, CA conjecture predicts that the complexity growth rate is zero at early time and then changes from negative infinite at a particular time~\cite{Lehner:2016vdi}. The investigation on the complexity between time-dependent TFD states in pure quantum field theory can give us evidence to judge which one of CV and CA conjectures is better.


\acknowledgments

I would like to thank Keun-Young Kim, Chao Niu, Yunlong Zhang, Yonghui Qi, Shaojiang Wang, Yusen An and Song He for valuable discussions. I also thank Rob. Myers and Shira Chapman for their correspondences. We also would like to thank workshop “Holography and Geometry of quantum criticality” at APCTP in Pohang of Korea,  where part of this work was done.

\appendix
\section{Transformation rule in different bases}\label{transF1}
In this appendix, I will give a simple explicit example about how to find the function form for a Finsler structure in the new basis $E'$ if $E'$ and the original basis are associated by transformation ${A^J}_I$ shown in Eq.~\eqref{E1toE2}.

Let's consider the case that the generator set $E=\{\hat{M}^1,\hat{M}^2,\cdots\}$. This generators set can generate an operators set $\mathcal{U}$. For any tangent operator $\hat{T}(t)$, we can decompose it as,
\begin{equation}\label{decompT1}
  \hat{T}(t)=Y_1(t)\hat{M}^1+Y_2(t)\hat{M}^2+\cdots\,.
\end{equation}
Let's assume that the function form of Finsler structure in this basis is given by $F_p$ form
\begin{equation}\label{FT1a}
  F[\hat{c}(t),Y_1(t),Y_2(t),\cdots]=p^I\parallel Y_I(t)\parallel=p^1\parallel Y_1(t)\parallel+p^2\parallel Y_2(t)\parallel+\cdots\,
\end{equation}
Here $\parallel\cdot\parallel$ is defined in Eq.~\eqref{twoF1}. In the new generators set $E'=\{\hat{M}'^1,\hat{M}'^2,\cdots\}$, the tangent operator can also be decomposed into following form,
\begin{equation}\label{decompT2}
  \hat{T}(t)=Y_1'(t)\hat{M}'^1+Y_2'(t)\hat{M}'^2+\cdots\,.
\end{equation}
In the new basis, the function form of Finsler structure can ba expressed by the function of coefficients $\{Y_1'(t), Y_2'(t), \cdots\}$.
The coefficients $\{Y'_1(t), Y'_2(t),\cdots\}$ and $\{Y_1(t), Y_2(t),\cdots\}$ are associated by Eq.~\eqref{tranforY}. If we require that  two function form give the same Finsler structure, then $F$ and $F'$ must satisfy the Eq.~\eqref{twoFs}, i.e.,
\begin{equation}\label{relFFs1}
  F'[\hat{c}(t),Y'_1(t),Y'_2(t),\cdots]=p^1\parallel Y_1(t)\parallel+p^2\parallel Y_2(t)\parallel+\cdots\,.
\end{equation}
Then we obtain following the transformation rule for function forms of Finsler structure under the basis transformation,
\begin{equation}\label{relFFs1}
  F'[\hat{c}(t),Y'_1(t),Y'_2(t),\cdots]=p^J\parallel {A^I}_JY'_I\parallel\neq p^I\parallel Y'_I\parallel\,.
\end{equation}
We see that after a basis transformation, the new function form of Finsler structure is different from form such as $F_p$ shown in Eq,~\eqref{fourFs}.

\section{Uniqueness of ladder operators for given Hamilton}\label{app1}
In this appendix, I will show that for given Hamilton $\hat{H}$ with discrete eigenvalues, there is a unique group of ladder operators which can satisfies the requirements in Eq.~\eqref{ladders}.
Assume that the state $|E_n,\vec{k}\rangle$ is the eigenstate corresponding to the eigenvalue $E_n$. Then we see that,
\begin{equation}\label{HEn1}
  \hat{H}=\sum_{n=0,\vec{k}}^{\infty}E_n|E_n,\vec{k}\rangle\langle E_n,\vec{k}|\,.
\end{equation}
The requirement
\begin{equation}\label{reqls1}
  \hat{l}_{\vec{k}}|E_n,\vec{k}\rangle=\alpha_{n,\vec{k}}|E_{n-1},\vec{k}\rangle\,
\end{equation}
shows that operator $\hat{l}_{\vec{k}}$ must have following form,
\begin{equation}\label{defl1}
  \hat{l}_{\vec{k}}=\sum_{n,m=0}^{\infty}\alpha_{n,\vec{k}}\delta_{m+1,n}|E_m,\vec{k}\rangle\langle E_n,\vec{k}|=\sum_{n=1}^{\infty}\alpha_{n,\vec{k}}|E_{n-1},\vec{k}\rangle\langle E_n,\vec{k}|
\end{equation}
Here $\alpha_n>0$ for $n>1$ and $\alpha_{0,\vec{k}}=0$. Then one can easy check that this operator satisfies following equation,
\begin{equation}\label{ladders1}
  \hat{l}^{\dagger}_{\vec{k}}|E_n,\vec{k}\rangle=\alpha_{n+1,\vec{k}}|E_{n+1},\vec{k}\rangle\,.
\end{equation}
After some algebras, we can obtain that,
\begin{equation}\label{lls2}
  \hat{l}_{\vec{k}}\hat{l}^\dagger_{\vec{k}'}=\sum_{n=0}^{\infty}\alpha_{n+1,\vec{k}}^2\delta_{{\vec{k}},{\vec{k}'}}|E_n,\vec{k}\rangle\langle E_n,\vec{k}'|
\end{equation}
and,
\begin{equation}\label{lls2}
  \hat{l}^\dagger_{\vec{k}'}\hat{l}_{\vec{k}}=\sum_{n=0}^{\infty}\alpha_{n,\vec{k}}^2\delta_{{\vec{k}},{\vec{k}'}}|E_{n},\vec{k}\rangle\langle E_{n},\vec{k}'|
\end{equation}
One can read that the generalized particle number density operator $\hat{N}'_{\vec{k}}:=\hat{l}^\dagger_{\vec{k}}\hat{l}_{\vec{k}}$ is commutative to Hamilton. The commutator $\hat{l}^\dagger$ and $\hat{l}$ then reads,
\begin{equation}\label{commuls1}
  [\hat{l}_{\vec{k}},\hat{l}^\dagger_{\vec{k}'}]=\sum_{n=0}^{\infty}(\alpha_{n+1,{\vec{k}}}^2-\alpha_{n,{\vec{k}}}^2)\delta_{{\vec{k}},{\vec{k}'}}|E_{n},{\vec{k}}\rangle\langle E_{n},{\vec{k}'}|
\end{equation}
If $\alpha_{n+1,{\vec{k}}}^2-\alpha_{n,{\vec{k}}}^2=1$ then we can see that $[\hat{l}_{\vec{k}},\hat{l}^\dagger_{\vec{k}'}]=\delta_{{\vec{k}},{\vec{k}'}}\hat{\mathbb{I}}$. Thus, we find that $\alpha_{n,{\vec{k}}}=\sqrt{n}$. Therefore, there is a unique operator,
\begin{equation}\label{defl2}
  \hat{l}_{\vec{k}}:=\sum_{n=1}^{\infty}\sqrt{n}|E_{n-1},\vec{k}\rangle\langle E_n,\vec{k}|
\end{equation}
can satisfy the requirements in Eq.~\eqref{ladders}. $\hat{l}_{\vec{k}}$ is a lowering operator which can transform the energy eigenstate into the lower level and $\hat{l}_{\vec{k}}^\dagger$ is a raiseing operator which can transform the energy eigenstate into the higher level. One can see that this raising/lowering operator returns to the creation/annihilation operator in free field theory. For free field, the Hamilton and ladder operators have a simple relationship,
\begin{equation}\label{HNrel1}
  \hat{H}=\sum_{\vec{k}}\omega_{\vec{k}}\hat{l}^\dagger_{\vec{k}}\hat{l}_{\vec{k}}+E_0\,,
\end{equation}
for momentum dependent function $\omega_{\vec{k}}$ and a zero-point energy $E_0$. However, for general cases that $E_n$ is not the linear function of $n$, the Hamilton $\hat{H}$ and ladder operators do not satisfy the Eq.~\eqref{HNrel1}. Thus, we see that
\begin{equation}\label{HNrel2}
  \hat{H}\neq\sum_{\vec{k}}\omega_{\vec{k}}\hat{l}^\dagger_{\vec{k}}\hat{l}_{\vec{k}}+E_0\,
\end{equation}
for general interacted system.

\section{Finding the complexity when $E=E^0$}\label{coherent1}
Under the general generators set $E$ defined in Eq.~\eqref{extendbasis}, it seems to be a subtle and high technical problem to find the complexity and give out an exact proof. It will bring us far away from the physical aspects if we focus on this point. However, if we reduce the generators set to a very small and simple case, it is possible to give a short and exact proof about how to find the complexity generated by this small generators set. By doing this, it is also a good example to show the basic idea to find the complexity and obtain some direct feelings about framework in the paper.

Let's restrict the generators set $E=E^0=\{\hat{a}^\dagger,\hat{a}, \hat{\mathbb{I}}\}$,  which forms a h(1) Lie algebra. The operators set can be given by three independent complex-valued constants $\alpha_\pm$ and $\alpha_0$ by,
\begin{equation}\label{Opset1}
  \mathcal{U}:=\left\{\hat{U}(\alpha_+,\alpha_-,\alpha_0)~\left|\forall \alpha_\pm,\alpha_0\in\mathbb{C}, \hat{U}(\alpha_+,\alpha_-,\alpha_0):=\exp(\alpha_+\hat{a}^\dagger+\alpha_-\hat{a}+\alpha_0\hat{\mathbb{I}})\right.\right\}
\end{equation}
Now let's compute the complexity of any operator in the operator set $\mathcal{U}$.
The group multiplication law in the H(1) group takes the form,
\begin{equation}\label{giveUi12}
  \hat{U}(\alpha_+,\alpha_-,\alpha_0)\hat{U}(\alpha'_+,\alpha'_-,\alpha'_0)=\hat{U}(\alpha_++\alpha'_+,\alpha_-+\alpha'_-,\alpha_0+\alpha'_0+\frac12(\alpha_+\alpha'_--\alpha_-\alpha'_+)\,.
\end{equation}
It is very useful when we compute the complexity for coherent states and the operators in $\mathcal{U}$.

To compute the complexity of an operator in $\mathcal{U}$, we have to find the minimal length connecting it and identity. Any curve starting from the identity can be given by three complex functions $Y_\pm$ and $Y_0$ in this way,
\begin{equation}\label{E0curve1}
  \hat{c}(s):=\overleftarrow{P}\exp\int_0^s\td x[Y_+(x)\hat{a}^\dagger+Y_-(x)\hat{a}+Y_0(x)\hat{\mathbb{I}}]\,.
\end{equation}
Assume curve $\hat{c}(s)$ can connect $\hat{U}(\alpha_+,\alpha_-,\alpha_0)$ and identity, then we see that,
\begin{equation}\label{relay1y2}
  \overleftarrow{P}\exp\int_0^1\td s[Y_+(s)\hat{a}^\dagger+Y_-(s)\hat{a}+Y_0(s)\hat{\mathbb{I}}]=\exp(\alpha_+\hat{a}^\dagger+\alpha_-\hat{a}+\alpha_0\hat{\mathbb{I}})\,.
\end{equation}
Let's first find the relationship between $\{Y_\pm(s),Y_0(s)\}$ and $\{\alpha_\pm,\alpha_0\}$. As $Y_+(s)\hat{a}^\dagger+Y_-(s)\hat{a}+Y_0(s)$ is not commutative to each other for different $s$, the time-order operator cannot be dropped.  We rewrite the time-order integration into the time-order product form,
\begin{equation}\label{realy1y2a}
  \overleftarrow{P}\exp\int_0^1\td s[Y_+(s)\hat{a}^\dagger+Y_-(s)\hat{a}+Y_0(s)\hat{\mathbb{I}}]=\overleftarrow{P}\prod_{n=0}^\infty\hat{g}_n
\end{equation}
with,
\begin{equation}\label{defGn1}
  \hat{g}_n:=\exp\{\Delta s[Y_+(s_n)\hat{a}^\dagger+Y_-(s_n)\hat{a}+Y_0(s_n)\hat{\mathbb{I}}]\}\,.
\end{equation}
Here $\Delta s\rightarrow0$ and $s_n=n\Delta s$. Now let's introduce series $\{b_n^{(\pm)},b_n^{(0)}\}$ and define $$\overleftarrow{P}\prod_{k=0}^n\hat{g}_k=\exp(b^{(+)}_n\hat{a}^\dagger+b^{(-)}_n\hat{a}+b^{(0)}_n).$$
Then we see that $\alpha_\pm=\lim_{n\rightarrow\infty}b^{(\pm)}_n$, and $\alpha_0=\lim_{n\rightarrow\infty}b^{(0)}_n$. On the other hand, we can find that,
\begin{equation}\label{anbncn1}
\begin{split}
  &\hat{g}_{n+1}\exp(b^{(+)}_n\hat{a}^\dagger+b^{(-)}_n\hat{a}+b^{(0)}_n\hat{\mathbb{I}})\\
  =&\exp\left\{[b^{(+)}_n+\Delta sY_+(s_{n+1})]\hat{a}^\dagger+[b^{(-)}_n+\Delta sY_-(s_{n+1})]\hat{a}+\right.\\
  &\left.b^{(0)}_n\hat{\mathbb{I}}+\Delta sY_0(s_{n+1})\hat{\mathbb{I}}+\frac{\hat{\mathbb{I}}\Delta s}2[Y_+(s_{n+1})b^{(-)}_n-b^{(+)}_nY_-(s_{n+1})]\right\}
  \end{split}
\end{equation}
Thus, there are following recursion relationships,
\begin{equation}\label{recursion1}
\begin{split}
  &b^{(+)}_{n+1}=b^{(+)}_n+\Delta sY_+(s_{n+1}),~~b^{(-)}_{n+1}=b^{(-)}_n+\Delta sY_-(s_{n+1}),\\
  &b^{(0)}_{n+1}=b^{(0)}_n+\Delta sY_0(s_{n+1})+\frac{\Delta s}2[Y_+(s_{n})b^{(-)}_n-Y_-(s_{n})b^{(+)}_n]+\mathcal{O}(\Delta s^2)
  \end{split}
\end{equation}
After dropping the higher order infinitesimal $\mathcal{O}(\Delta s^2)$, we find that the solutions read,
\begin{equation}\label{solutionabc}
\begin{split}
  &b^{(+)}_n=\sum_{k=0}^n\Delta sY_+(s_{n}),~~b^{(-)}_n=\sum_{k=0}^n\Delta sY_-(s_{n}),\\
  &b^{(0)}_n=\sum_{k=0}^n\Delta s\{Y_0(s_{n})+\frac12[Y_+(s_n)\sum_{k=0}^nY_-(s_n)-Y_-(s_n)\sum_{k=0}^nY_+(s_n)]\}
  \end{split}
\end{equation}
Now taking $n\rightarrow\infty$ and converting them into the continuous form, we find following relationship between $\{Y_\pm(s),Y_0(s)\}$ and $\{\alpha_\pm,\alpha_0\}$,
\begin{equation}\label{relalphaY}
  \alpha_\pm=\int_0^1\td sY_\pm(s),~~\alpha_0=\frac12\int_0^1\td s\left[2Y_0(s)+Y_+(s)\int_0^s\td\tilde{s}Y_-(\tilde{s})-Y_-(s)\int_0^s\td\tilde{s}Y_+(\tilde{s})\right]\,.
\end{equation}
Thus, we obtain following restricted optimization problem to find the complexity of $\hat{U}(\alpha_+,\alpha_-,\alpha_0)$,
\begin{equation}\label{findCUabc}
  \mathcal{C}[U(\alpha_+,\alpha_-,\alpha_0)]=\min\left\{\left.\int_0^1\td s[\parallel Y_+(s)\parallel+\parallel Y_-(s)\parallel]\right|~\int_0^1\td sY_\pm(s)=\alpha_\pm\right\}
\end{equation}
This restricted optimization problem can be solved easy if we note the relationship $\parallel x+y\parallel\leq\parallel x\parallel+\parallel y\parallel$ which leads that $\int_0^1\td s\parallel Y_\pm(s)\parallel\geq\parallel\int_0^1\td s Y_\pm(s)\parallel$ and
\begin{equation}\label{solY1Y2}
  \int_0^1\td s[\parallel Y_+(s)\parallel+\parallel Y_-(s)\parallel]\geq\parallel \int_0^1\td sY_+(s)\parallel+\parallel \int_0^1\td sY_-(s)\parallel=\parallel \alpha_+\parallel+\parallel \alpha_-\parallel\,.
\end{equation}
Thus the complexity of $U(\alpha_+,\alpha_-,\alpha_0)$ is,
\begin{equation}\label{solCU}
  \mathcal{C}[U(\alpha_+,\alpha_-,\alpha_0)]=\parallel \alpha_+\parallel+\parallel \alpha_-\parallel\,.
\end{equation}

Now let's give the method to find the complexity between two coherent states. Let's take two different coherent states $|\text{coh}(b_1)\rangle$ and $|\text{coh}(b_2)\rangle$. Then we have,
\begin{equation}\label{definchob1b2}
  |\text{coh}(b_i)\rangle\sim\exp(b_i\hat{a}^\dagger)|0\rangle,~~~~i=1,2\,.
\end{equation}
To find the complexity between this two states, we have first to find all the operators in $\mathcal{U}$ which can satisfy $|\text{coh}(b_2)\sim\hat{U}|\text{coh}(b_1)\rangle$. For any operator parameterized by $\alpha_\pm$ and $\alpha_0$, we have following relationship
\begin{equation}\label{relUb2}
\begin{split}
  \hat{U}(\alpha_+,\alpha_-,\alpha_0)|\text{coh}(b_1)\rangle&\sim\exp(\alpha_+\hat{a}^\dagger+\alpha_-\hat{a})\exp(b_1\hat{a}^\dagger)|0\rangle\\
  &=\exp\left[(\alpha_++b_1)\hat{a}^\dagger+\alpha_-\hat{a}-\frac{\alpha_-b_1}2\right]|0\rangle\\
  &\sim\exp\left[(\alpha_++b_1)\hat{a}^\dagger\right]|0\rangle\,.
  \end{split}
\end{equation}
We see that $|\text{coh}(b_2)\rangle\sim\hat{U}|\text{coh}(b_1)\rangle$ if and only if $\alpha_+=b_2-b_1$. There are infinite different operators which can convert $|\text{coh}(b_1)\rangle$ to $|\text{coh}(b_1)\rangle$. The complexity between any two coherent states is,
\begin{equation}\label{Ccohb1b2}
\begin{split}
  \mathcal{C}[|\text{coh}(b_2)\rangle, |\text{coh}(b_1)\rangle]&=\min\{\parallel \alpha_+\parallel+\parallel \alpha_-\parallel|~\forall\alpha_-\in\mathbb{C},~\text{and}~\alpha_+=b_2-b_1\}\\
  &=\parallel b_2-b_1\parallel\,.
  \end{split}
\end{equation}

\section{Finding the complexity in a larger generators set for TFD states}\label{app2}
In the subsection~\ref{check2bTFD}, we have chosen the generators set to be $E=\{\hat{L}^{(\vec{k})}_+, \hat{\mathbb{I}}|~\forall \vec{k}\in\mathbb{R}^{d-1}\}$ to compute the complexity. In this appendix, I will extend it to include the Casimir operators given by Eq.~\eqref{Casimir1}, i.e,
\begin{equation}\label{extE0TFD}
  E=\{\hat{L}^{(\vec{k})}_+, \hat{C}^{(\vec{k})},\hat{\mathbb{I}}|~\forall \vec{k}\in\mathbb{R}^{d-1}\}\,.
\end{equation}
We see that $\hat{L}^{(\vec{k})}_+$ contains 2 creation operators and $\hat{C}^{(\vec{k})}$ contains 4 creation/anihilation operators. By definition, the weight for generator $\hat{L}^{(\vec{k})}_+$ is just 2 but the weight for $\hat{C}^{(\vec{k})}$ is not 4. According to Eq.~\eqref{Casimir1}, we can see that,
\begin{equation}\label{Casimir2}
\begin{split}
  \hat{C}^{(\vec{k})}&=\hat{L}_0^{(\vec{k})2}-\hat{L}_0^{(\vec{k})}-\hat{L}_+^{(\vec{k})}\hat{L}_-^{(\vec{k})}\\
  &=\frac14(\hat{a}^{R\dagger2}_{\vec{k}}\hat{a}^{R2}_{\vec{k}}+\hat{a}^{L\dagger2}_{\vec{k}}\hat{a}^{L2}_{\vec{k}})+\frac{\hat{N}_{\vec{k}_i}^{R}\hat{N}_{\vec{k}_i}^{L}}{2}+\frac14(\hat{N}_{\vec{k}_i}^{R}+\hat{N}_{\vec{k}_i}^{L}) -\hat{L}_+^{(\vec{k})}\hat{L}_-^{(\vec{k})}-\frac{\hat{\mathbb{I}}}4
  \end{split}
\end{equation}
Here $\hat{N}_{\vec{k}_i}^{R}$ and $\hat{N}_{\vec{k}_i}^{L}$ are particle density operators. The last line of Eq.~\eqref{Casimir2} has been rewritten as the summation of some normal order polynomials of creation and annihilation operators. Using the transformation rule~\eqref{relFFs1},  we see that the weight for $\hat{C}^{(\vec{k})}$ should be $(4+4)/4+4/2+(2+2)/4+4=9$.
As $E$ forms an Abelian Lie algebra, any operator generated by $E$ can be parameterized by three complex functions $f_1(\vec{k}), f_2(\vec{k})$ and $ f_0(\vec{k})$ in following way,
\begin{equation}\label{Uf1f2s}
  \hat{U}(f_0, f_1,f_2):=\exp\int[f_1(\vec{k})\hat{L}^{(\vec{k})}_++f_2(\vec{k})\hat{C}^{(\vec{k})}+f_0(\vec{k})\hat{\mathbb{I}}]\td^{d-1}k\,.
\end{equation}
Then the operators set $\mathcal{U}_E$ is given by,
\begin{equation}\label{extUset1}
  \mathcal{U}_E:=\{\hat{U}(f_0, f_1,f_2)~|~\forall f_0, f_1,f_2:\mathbb{R}^{d-1}\mapsto\mathbb{C}\}\,.
\end{equation}
The operators set $\mathcal{U}_E$ forms an  Abelian Lie group with the group multiplication law,
\begin{equation}\label{extTFDlaw1}
  \hat{U}(f_0, f_1,f_2)\hat{U}(\tilde{f}_0, \tilde{f}_1,\tilde{f}_2)=\hat{U}(f_0+\tilde{f}_0, f_1+\tilde{f}_1,f_2+\tilde{f}_2)
\end{equation}
Any curve staring from the identity can be given by three complex function $y_1(t,\vec{k}), y_2(t,\vec{k})$ and $ y_0(t,\vec{k})$ in this way,
\begin{equation}\label{Uf1f3s}
  \hat{c}(s):=\overleftarrow{\mathcal{P}}\exp\int_0^s\td t\int\td^{d-1}k[y_1(t,\vec{k})\hat{L}^{(\vec{k})}_++y_2(t,\vec{k})\hat{C}^{(\vec{k})}+y_0(t,\vec{k})\hat{\mathbb{I}}]\,.
\end{equation}
As $\mathcal{U}_E$ forms an  Abelian Lie group, the time order operator can be dropped. The condition that $\hat{c}(1)=\hat{U}(f_0, f_1,f_2)$ leads to,
\begin{equation}\label{extTFDcond}
  \int_0^1y_i(t,\vec{k})\td t=f_i(k),~~~i=0,1,2\,.
\end{equation}
Then computing the complexity for operator $\hat{U}(f_0, f_1,f_2)$ becomes following optimization problem under the restrictions~\eqref{extTFDcond},
\begin{equation}\label{extCU1}
  \mathcal{C}[\hat{U}(f_0, f_1,f_2)]=\frac{\text{Vol}}{(2\pi)^{d-1}}\min\{\int_0^1\td t\int\td k^{d-1}[2\parallel y_1(t, \vec{k})\parallel+9\parallel y_2(t, \vec{k})\parallel]\}\,.
\end{equation}
Similar to Eq.~\eqref{solY1Y2}, we can see that,
\begin{equation}\label{extCU2}
  \mathcal{C}[\hat{U}(f_0, f_1,f_2)]=\frac{\text{Vol}}{(2\pi)^{d-1}}\left[2\parallel\int\td k^{d-1} f_1(\vec{k})\parallel+9\parallel \int\td k^{d-1}f_2(\vec{k})\parallel\right]\,.
\end{equation}
For the case that $f_0=f_2=0$ and $f_1(\vec{k})=\lambda(\vec{k})$, we can see that the complexity is,
\begin{equation}\label{extCU2}
  \mathcal{C}[\hat{U}(0, \lambda,0)]=\frac{2\text{Vol}}{(2\pi)^{d-1}}\parallel\int\td k^{d-1} \lambda(\vec{k})\parallel\,,
\end{equation}
which is the same as Eq.~\eqref{ComFTDU1}. We see that the complexity is the same even we extend the generators set from Eq.~\eqref{TFDgenerE} into Eq.~\eqref{extE0TFD}.

These steps can be generalized into a more larger generator set,
\begin{equation}\label{extE0TFD2}
  E=\{\hat{L}^{(\vec{k})n}_+, \hat{C}^{(\vec{k})n}, \hat{\mathbb{I}}|~\forall \vec{k}\in\mathbb{R}^{d-1}, \forall n\in\mathbb{N}^+\}\,.
\end{equation}
This generators set contains infinite different polynomials of creation and annihilation operators but does not contain polynomials of creation and annihilation operators between different momentums. One can see that it can still generate an Abelian group. It is easy to see that the complexity of $\exp\int\td^dk\lambda(\vec{k})\hat{L}^{(\vec{k})}_+$ in this generators set is still given by Eq.~\eqref{ComFTDU1}.

\section{The basic inequality for normal $\parallel\cdot\parallel$}\label{basicineq}
In this appendix, the proof will be given to show a basic inequality for the normal $\parallel\cdot\parallel$ defined in Eq.~\eqref{twoF1}. For any two complex constants $x$ and $y$, I will show that,
\begin{equation}\label{basicineq1}
  \parallel x\parallel+\parallel y\parallel\geq\parallel x+y\parallel\,.
\end{equation}
When $x$ and $y$ are both real-valued, this is just the basic inequality about the absolute value. Let's pay attention to the case that one of them has nonzero imaginary part.  In order to prove the inequality \eqref{basicineq1}, let's refer to following lemma (see the chapter 1.2 in Ref.~\cite{9810245319}):\\
\textit{\textbf{Lemma}}: Suppose $V$ to be a vector space. For any nonnegative function $H: V\mapsto [0,\infty)$, if it satisfies following three conditions:\\
(M1)~$H$ is $C^\infty$ on $V\setminus\{0\}$;\\
(M2)~$H(\lambda x)=\lambda H(x)$ for $\forall x\in V$ and $\lambda>0$;\\
(M3)~$\forall x\in V\setminus\{0\}$ and $\forall y_1,y_2\in V$, following symmetric bilinear form
\begin{equation}\label{basicg1}
  \left.\frac{\partial^2}{\partial u\partial v}H^2(x+uy_1+vy_2)\right|_{u=v=0}\,
\end{equation}
is positive, then we have following triangle inequality,
\begin{equation}\label{Finstriang}
  H(x)+H(y)\geq H(x+y),~~\forall x,y\in V
\end{equation}
and the equality can be achieved if and only if $\exists \lambda\geq0$ such that $x=\lambda y$.

The normal $\parallel\cdot\parallel$ in Eq.~\eqref{twoF1} is defined in the complex number, which can be regarded as the function defined in $V=\mathbb{R}^2$. We can defined the function $H$ as,
\begin{equation}\label{definH1}
  H(a,b):=\parallel a+i b\parallel=\rho(a,b)(|\cos\theta(a,b)|+|\theta(a,b)||\sin\theta(a,b)|)\,
\end{equation}
with $\rho(a,b)=\sqrt{a^2+b^2}$. According to subsection~\ref{F-struc}, the function $\theta(a,b)$ is given in following way. If $a+ib$ is a constant, then $$\theta=\theta_0:=\arccos(a/\sqrt{a^2+b^2}).$$ If $a+ib$ is the function of $s$ so that $a+ib$ forms a smooth curve in the complex plan then $$\theta(s)=\theta_0+n(s)\pi$$ for an integer $n(s)$ which is determined by the requirements that $n(0)=0$ and $n(s)$ can make the function $\theta(s)$ to be $C^\infty$.

It is easy to see that $H$ defined in Eq.~\eqref{definH1} satisfies the conditions (M1) and (M2). Now let's prove that $H(a,b)$ also satisfies the condition (M3). The requirement (M3) is equivalent to that matrix $$\left[\begin{matrix}
     \frac{\partial^2 H^2}{\partial a^2}, & \frac{\partial^2 H^2}{\partial a\partial b} \\
     \frac{\partial^2 H^2}{\partial a\partial b}, & \frac{\partial^2 H^2}{\partial b^2}
\end{matrix}\right]$$ is positive definite matrix,
which is yield following requirements,
\begin{equation}\label{M3eq1}
  \frac{\partial^2 H^2}{\partial a^2}>0,\frac{\partial^2 H^2}{\partial a^2}\frac{\partial^2 H^2}{\partial b^2}-\left(\frac{\partial^2 H^2}{\partial a\partial b}\right)^2>0\,.
\end{equation}
%
One can check by direct computation that the function $H(a,b)$ defined in Eq.~\eqref{definH1} satisfies the requirments in Eq~\eqref{M3eq1}.
Thus the inequality \eqref{basicineq1} is true.

\bibliographystyle{JHEP}
\bibliography{ref}

\end{document}